\documentclass[12pt]{iopart}

\usepackage{graphicx}
\usepackage{multirow}
\usepackage{enumerate}
\usepackage{multicol}
\usepackage{color}

\begin{document}

\title[Modelling of conventional and Super-X divertor configurations of MAST Upgrade]{Investigation of conventional and Super-X divertor configurations of MAST Upgrade using SOLPS}

\author{E Havl\'i\v ckov\'a$^1$, W Fundamenski$^2$, M Wischmeier$^3$, G Fishpool$^1$, A W Morris$^1$}

\address{$^1$ EURATOM/CCFE Fusion Association, Culham Science Centre, Abingdon, Oxon, OX14 3DB, United Kingdom}
\address{$^2$ Imperial College of Science, Technology and Medicine, London, UK}
\address{$^3$ Max-Planck Institut f\"ur Plasmaphysik, EURATOM Association, Garching, Germany} 

\ead{eva.havlickova@ccfe.ac.uk}

\begin{abstract}
One of the first studies of MAST Upgrade divertor configurations with SOLPS5.0 are presented. We focus on understanding main prospects associated with the novel geometry of the Super-X divertor (SXD). This includes a discussion of the effect of magnetic flux expansion and volumetric power losses on the reduction of target power loads, the effect of divertor geometry on the divertor closure and distribution of neutral species and radiation in the divertor, the role of the connection length in broadening the target wetted area. A comparison in conditions typical for MAST inter-ELM H-mode plasmas confirms improved performance of the Super-X topology resulting in significantly better divertor closure with respect to neutrals (the atomic flux from the target increased by a factor of 6, but the atomic flux from the divertor to the upper SOL reduced by a factor of 2), increased radiation volume and increased total power loss (a factor of 2) and a reduction of target power loads through both magnetic flux expansion and larger volumetric power loss in the divertor (a factor of 5--10 in attached plasmas). The reduction of the target power load by SXD further increases with collisionality (high density or detached regimes) thanks to larger importance of volumetric power losses. It is found that a cold divertor plasma leads to stronger parallel temperature gradients in the SOL which drive more parallel heat flux, meaning that the effectiveness of perpendicular transport in spreading the power at the target can be reduced, and this needs to be taken into account in any optimisation.

\end{abstract}

\maketitle

\section{Introduction} 
Advanced divertor geometries such as Snowflake or Super-X divertors \cite{Ryutov,Valanju} are presently investigated as alternative configurations for future reactors, with an effort to provide solutions for power exhaust issues. The major aim is to effectively reduce power loads to divertor targets which could, in large devices, exceed a tolerable maximum set by material limits. In this paper, we study the effects of the Super-X divertor which will be installed on the MAST tokamak \cite{SXDMast}. Another expected benefit of such divertor is an improved closure, hence higher neutral pressure in the divertor, but limited penetration of fuelling species neutrals and impurities into the region above the X-point. 

2D SOL transport codes such as SOLPS \cite{Schneider} have proven to be helpful in quantifying the effects of geometry on the divertor performance and have been used to support the divertor design, e.g. on ITER \cite{ITER}. On the MAST tokamak, recent SOLPS5.0 simulations focused on optimization of the baffle positioning with respect to retaining the divertor closure, but avoiding large flux on the baffle surfaces \cite{John}. The effect of the Super-X divertor was preliminarily investigated also in \cite{Lisgo,Eva1} by SOLPS5.0 and in \cite{Eva2} by a 1D code SOLF1D. Some of the latest results also concentrate on drift effects \cite{Vladimir1,Vladimir2}.

Here we focus on the overall impact of the Super-X topology on the divertor performance in attached plasmas without impurity puffing. Simulations are based on two configurations proposed for MAST Upgrade (MAST-U): (i) a short-legged divertor (CD: the conventional divertor) which is closer to divertors in present machines (section 3), (ii) a long-leg divertor (SXD: the Super-X divertor) with more complex magnetic topology employing additional poloidal coils in order to increase the connection length $L_{\parallel}$ and expand the plasma to larger radius (section 4). 
A potential to optimize the effect of SXD in high density regimes and seeded plasmas will be discussed at the next stage. 

Beside a direct comparison of the two configurations for inter-ELM H-mode plasmas, the sensitivity of results to various factors has been studied. This includes scans for both physics parameters (such as input power to the SOL, density, radial transport coefficients, sputtering, gas puff location) and simulation parameters (e.g. fluid versus kinetic neutrals, heat flux limiters). As a baseline for MAST-U predictions, SOLPS5.0 results for MAST discharges, both L-mode and H-mode, are currently being validated against experiment as a separate work. With regards to limits of the codes as SOLPS and common problems with the interpretation of experimental results, the paper focuses on a comparative study of CD and SXD, rather than trying to quantitatively reproduce a MAST experiment. The study presented here is extended in \cite{Eva3} from low density attached conditions to high density and detached regimes.

\section{Simulation of the conventional divertor}
\label{sec_cd}

\subsection{Simulation setup}

\begin{figure}[!h]
\centerline{\scalebox{0.45}{\includegraphics[clip]{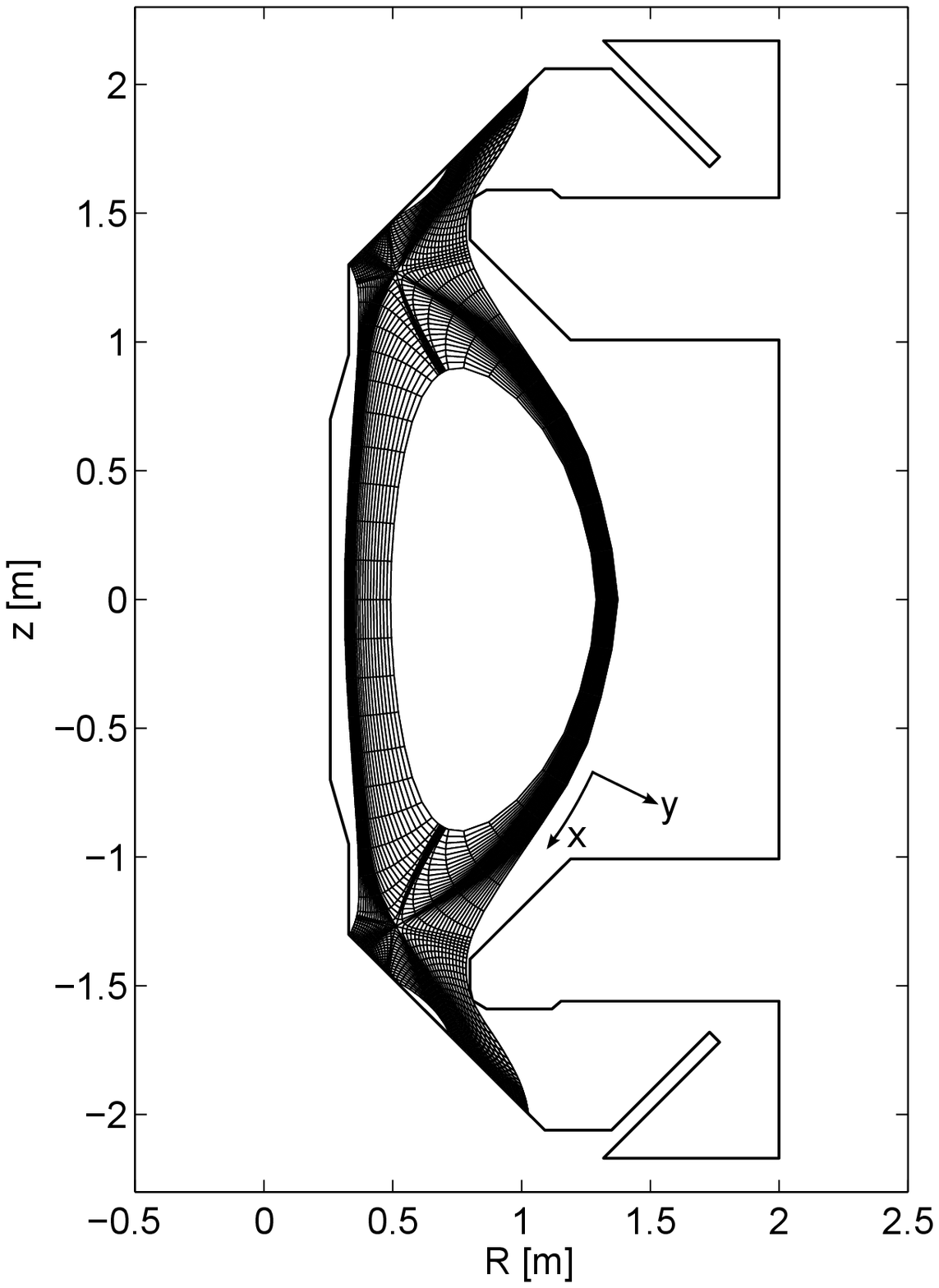}}
\scalebox{0.268}{\includegraphics[clip]{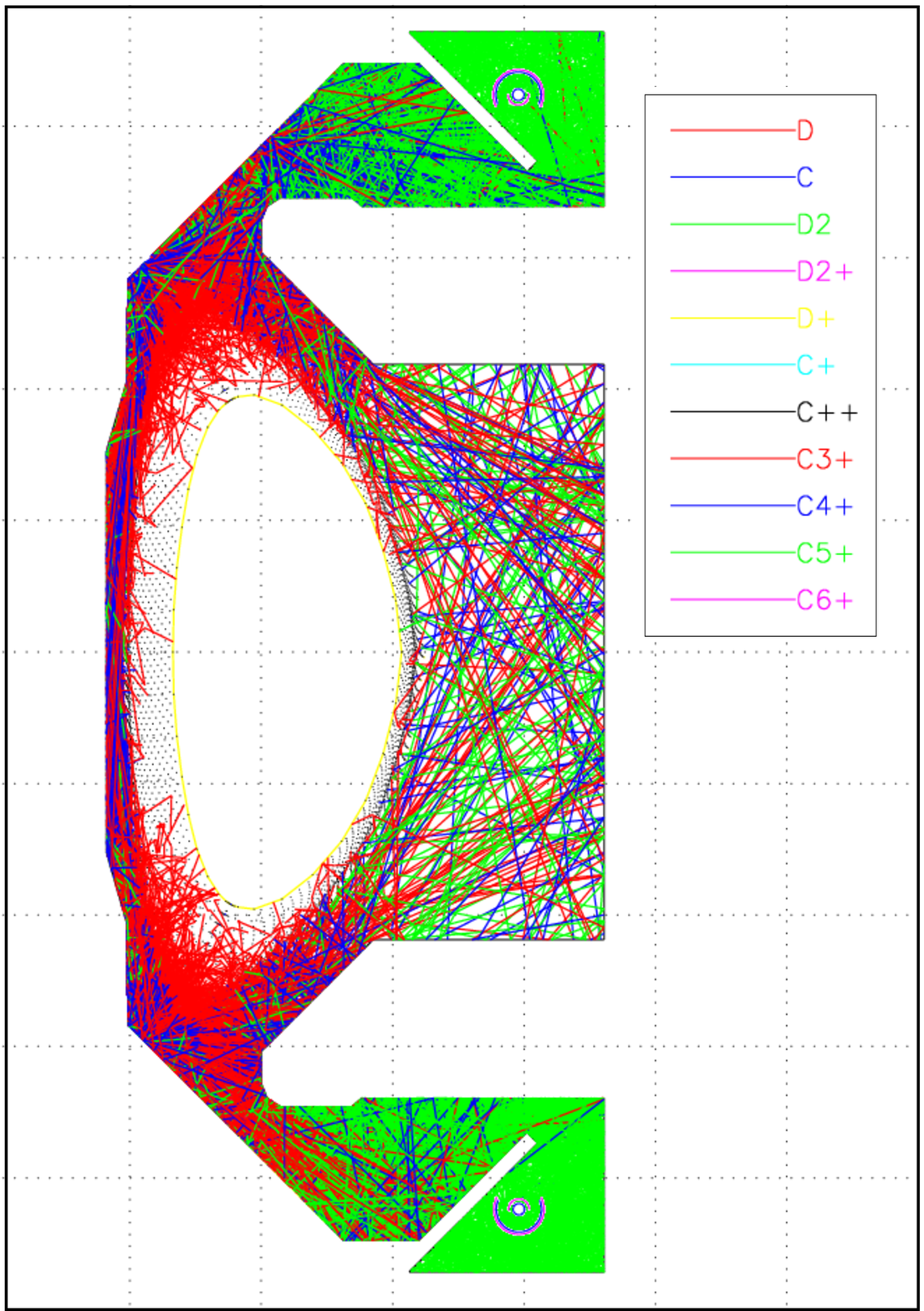}}
}
\caption{On the left, a SOLPS grid for the conventional divertor geometry (limited by the contact point with the baffle). Both Cartesian coordinates ($R$,$z$) and magnetic field coordinates -- poloidal and radial ($x$,$y$) are defined. On the right, the domain for EIRENE simulation.} \label{fig_grid}
\end{figure}

The divertor design for MAST-U offers large flexibility in terms of magnetic topology. One of the possible configurations is a conventional divertor shown in Fig. \ref{fig_grid}. SOLPS5.0 including EIRENE is used to study inter-ELM H-mode plasmas in this magnetic configuration with parameters typical for current MAST H-mode discharges. This section is to give a summary of main results for the conventional divertor for the purpose of comparison with results for the Super-X magnetic configuration in the second part of the paper. 

\begin{figure}[!h]
\centerline{\scalebox{0.65}{\includegraphics[clip]{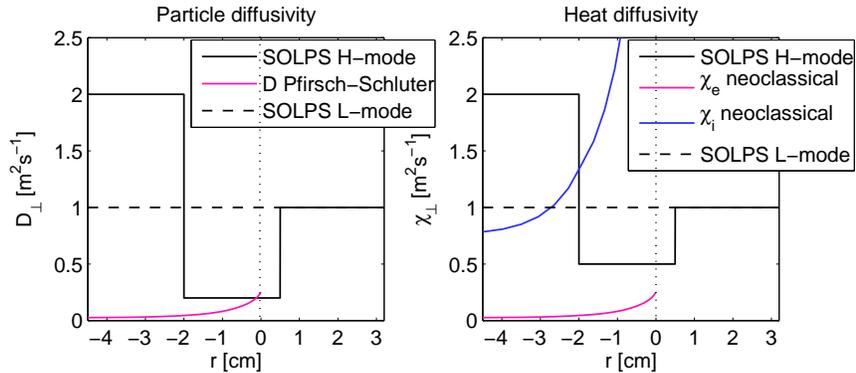}}
}
\caption{Radial diffusivities assumed in the code are compared with neoclassical Pfirsch-Schl\"uter values calculated from SOLPS profiles of the density and temperature for the case (a) according to expressions given in \cite{Wesson}. The diffusion coefficients are plotted as functions of the radial coordinate at the midplane defined as $r\equiv R-R_{\rm sep}$ with the separatrix at $r=0$.} \label{fig_dchi}
\end{figure}

We define a case (a) with $P_{\rm inp}=1.7$ MW (the total input power to the grid), $n_{\rm core}=2.8\times 10^{19}$ m$^{-3}$ (the density at the inner boundary of the grid).
The temperature and density drop across the pedestal is imposed by a transport barrier in perpendicular diffusivities (Fig. \ref{fig_dchi}) which are otherwise poloidally uniform. These diffusivities are based on \cite{Vladimir2} and an earlier benchmarking of SOLPS with experiment \cite{Vladimir3}.

As the SOL collisionality and the SOL width are important parameters of the SOL, we analyze three additional simulations defined in Tab. \ref{table_runs}: case (b) with larger radial transport assuming $D_{\perp}=1$ m$^2$s$^{-1}$ and $\chi_{\perp}=1$ m$^2$s$^{-1}$ (shown in Fig. \ref{fig_dchi} as dashed line), case (c) with twice larger $n_{\rm core}$, case (d) with twice lower $P_{\rm inp}$. While the case (d) gives SOL temperatures comparable to those observed in the current MAST experiment, the case (a) allows for larger heating power available in MAST-U. The case (b) will be referred to as L-mode (here meaning wider SOL, noting that L-mode in MAST is usually a low power Ohmic plasma, below 1.7 MW).

\begin{table}[!h]
\begin{center}
\begin{tabular}{lrrrr}
    & (a) & (b) & (c) & (d) \\ 
\hline
$n_{\rm core}$ [10$^{19}$ m$^{-3}$] & 2.8 & 2.8 & 5.6 & 2.8 \\
$P_{\rm inp}$ [MW] & 1.7 & 1.7 & 1.7 & 0.85 \\
$D_{\perp}$ [m$^{2}$s$^{-1}$] & 0.2 & 1.0 & 0.2 & 0.2 \\
$\chi_{\perp}$ [m$^{2}$s$^{-1}$] & 0.5 & 1.0 & 0.5 & 0.5 \\
\hline
\end{tabular}
\caption{Definition of four SOLPS simulations considered in the paper.}
\label{table_runs}
\end{center}
\end{table}

Apart from deuterium plasma species, sputtered carbon impurity is taken into account, assuming both physical and chemical sputtering, and the chemical sputtering yield of 1\%. The effect of larger sputtering of 3\% will be explored as well. The particle source is from the core and the flux crossing the separatrix is found to satisfy the given $n_{\rm core}$ mimicking the particle fuelling from a gas puff.
The source is balanced by a flux on cryopumps which are located behind the outer targets (Fig. \ref{fig_grid} right). The active pumping element of radius 2 cm gives the pumping speed of 42 m$^3$/s corresponding to a recycling coefficient $R=0.9$ on this element (compared to 11 m$^3$/s in the current MAST experiment). Concerning parallel transport coefficients, the viscosity limiter $\beta=0.5$ is used in all simulations following \cite{David,Wojtek}. The value of heat flux limiters is more questionable as it is more sensitive to the collisionality. Due to the lack of systematic study of kinetic corrections and their parametrization with plasma parameters, the heat flux here is prescribed by the classical Spitzer-H\"arm expression. Note that the simulation is toroidally symmetric and smooth targets (no imbrication) are assumed.

\subsection{Plasma parameters}

\begin{figure}[!h]
\centerline{\scalebox{0.68}{\includegraphics[clip]{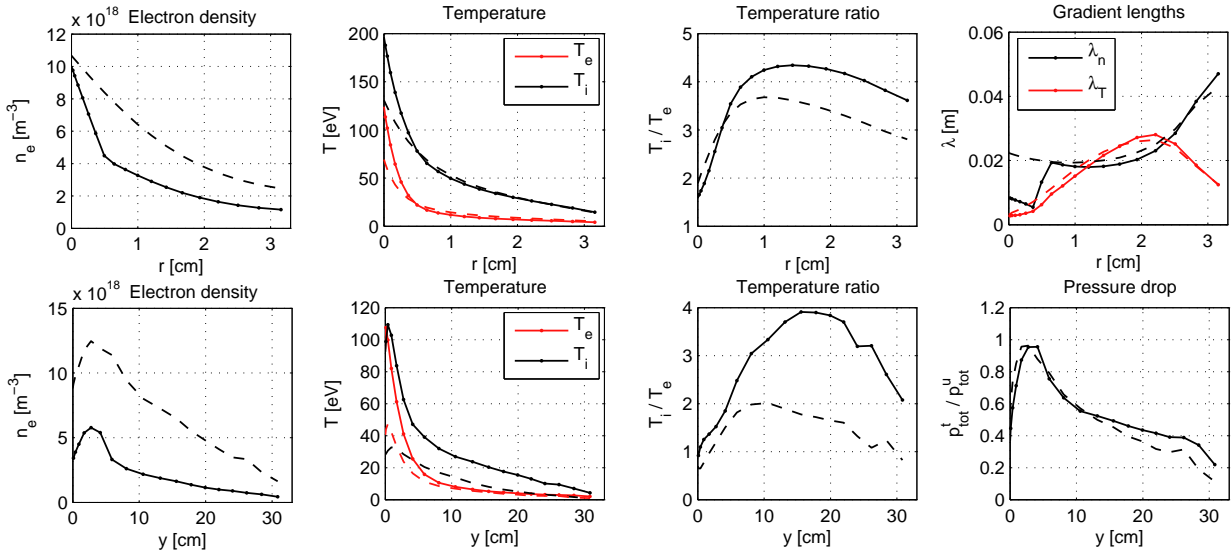}}
}
\caption{The CD configuration -- (top) midplane radial profiles of (i) the electron density, (ii) the electron and ion temperature, (iii) the ratio of the temperatures, (iv) the gradient lengths of the density and temperature. (bottom) Target radial profiles of (i) the electron density, (ii) the electron and ion temperature, (iii) the ratio of the temperatures, (iv) the pressure drop along the SOL measured as the ratio of the target and upstream total pressure $p_{\rm tot}=m_{\rm i}n_{\rm e}u_{\parallel}^2+n_{\rm e}k(T_{\rm e}+T_{\rm i})$. Here $y$ is the coordinate aligned with the target. The case (a) is shown as a solid line, while the case (b) is shown as a dashed line.} \label{fig_radial}
\end{figure}

The first subjects of comparison between CD and SXD are densities and temperatures in the divertor. 
Radial profiles of plasma parameters at the outer midplane and outer target are shown in Fig. \ref{fig_radial} for the cases (a) and (b) in CD. For given $n_{\rm core}$ and $P_{\rm inp}$, the density and collisionality in the SOL is larger in the case (b) as the result of larger $D_{\perp}$, while the temperatures at the midplane are lower in the near SOL, as well as in the divertor. 

MAST Upgrade will have a capability to measure the $T_{\rm i}/T_{\rm e}$ ratio using the retarding filed energy analyser, see e.g. \cite{Sarah}.
In the simulation, the temperature ratios of $T_{\rm i}/T_{\rm e}\approx 2$ are found at the midplane separatrix and go up to approximately 4 across the SOL. At the target, the temperature ratio ranges between 1--4 for the H-mode case (a), while we find approximately $T_{\rm i}\approx T_{\rm e}$ to $T_{\rm i}\approx 2T_{\rm e}$ for the L-mode case (b) as the result of stronger energy coupling between electrons and ions at higher collisionality.

By modifying the divertor geometry, one can achieve a broadening of the SOL (by effects of magnetic flux expansion and longer connection length), hence a reduction of the peak fluxes. The radial SOL width can be charactezied by gradient lengths which are also shown in Fig. \ref{fig_radial} and will be compared with SXD in section \ref{sec_par_sxd}. The midplane gradient lengths from Fig. \ref{fig_radial} ($\lambda_n=n_{\rm e}/|\nabla_r n_{\rm e}|$ and $\lambda_T=T_{\rm e}/|\nabla_r T_{\rm e}|$) range from $\lambda_n^{\rm u}\approx 1$ cm and $\lambda_T^{\rm u}\approx 0.5$ cm in the near SOL to $\lambda_n^{\rm u}\approx 2-4$ cm and $\lambda_T^{\rm u}\approx 2-3$ cm in the far SOL (index u as upstream). 
The SOL width in the upstream SOL and at the target (defining the target wetted area) and the peaking of target particle and energy fluxes are established as the result of competition between radial and parallel transport. Laminar transport codes as SOLPS are, however, unable to model radial transport self-consistently and therefore the gradient lengths in Fig. \ref{fig_radial} are in principle given by the assumed radial transport coefficients.

\begin{figure}[!h]
\centerline{\scalebox{0.68}{\includegraphics[clip]{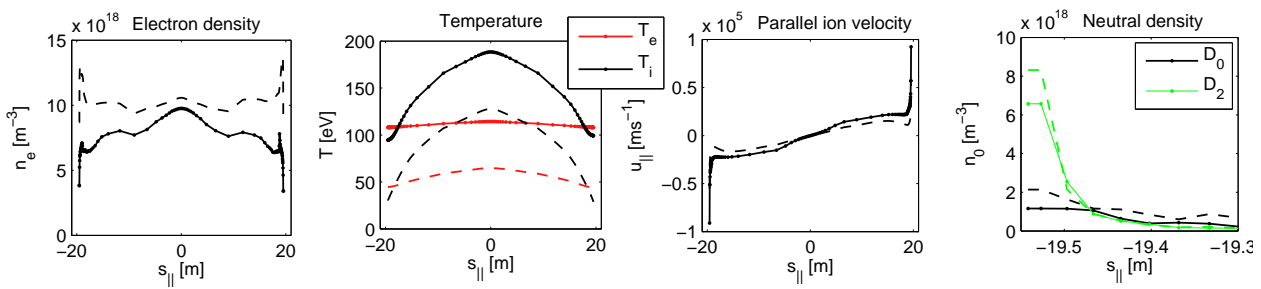}}
}
\caption{The CD configuration -- parallel profiles of (i) the electron density, (ii) the parallel ion velocity, (iii) the electron and ion temperature and (iv) the density of neutral species for the case (a) (solid) and (b) (dashed).} \label{fig_parallel}
\end{figure}

Parallel profiles in Fig. \ref{fig_parallel} (at the separatrix between the outer upper and lower targets) show sheath-limited SOL in the H-mode case (a) with flat $T_{\rm e}$ and low $n_{\rm e}$ at the target. With the peak target temperature of $T_{\rm e}\approx 100$ eV, the plasma is well attached and no pressure drop is observed along the field line (Fig \ref{fig_radial}). In the L-mode case (b), the plasma temperature at the target is approximately $T_{\rm e}\approx 40$ eV, a small $T_{\rm e}$ gradient along the SOL is developed (smaller parallel heat conductivity $\kappa_{\rm e}\propto T_{\rm e}^{5/2}$, larger $\nabla_{\parallel}T_{\rm e}$) and stronger collisional cooling in the divertor (thanks to larger plasma and neutral densities and smaller temperatures) steepens $\nabla_{\parallel}T_{\rm e,i}$ further. The plasma remains attached at the given $P_{\rm inp}$. The upstream collisionality is calculated in Tab. \ref{table1}, where the higher density and lower power cases (c) and (d) are included, all displaying low collisionality below 10, for which a sheath-limited SOL is expected and all with attached plasma with target temperatures above 20 eV. Note that the upstream and target parameters in Tab. \ref{table1} for CD are similar to those found in simulations with the current MAST divertor. 

\begin{table}[!h]
\begin{center}
\begin{tabular}{lrrrrrrrrr}
    & $L_{\parallel}$ & $n_{\rm e}^{\rm u}$ & $T_{\rm e}^{\rm u}$ & $T_{\rm i}^{\rm u}$ & $n_{\rm e}^{\rm t}$ & $T_{\rm e}^{\rm t}$ & $T_{\rm i}^{\rm t}$ & $\lambda_{\rm ei}$ & $\nu_{\rm e}$ \\   
    & [m] & [10$^{19}$ m$^{-3}$] & [eV] & [eV] & [10$^{19}$ m$^{-3}$] &  [eV] &  [eV] & [m] &\\  
\hline
(a) & 20 & 1.0 &  114 & 188 & 0.6 & 108 & 110 & 15.9 & 1.2 \\
(b) & 20 & 1.1 & 65 & 128 & 1.3 & 47 & 33 & 4.8 & 4.1 \\
(c) & 20 & 1.7 & 68 & 130 & 2.3 & 44 & 24 & 3.3 & 5.9 \\
(d) & 20 & 1.0 & 62 & 111 & 0.8 & 48 & 33 & 4.8 & 4.1 \\
\hline
\end{tabular}
\caption{The CD configuration -- the collisionality of the SOL plasma $\nu_{\rm e}$ and the mean free path $\lambda_{\rm ei}$ calculated as $\nu_{\rm e}=L_{||}/\lambda_{ei}$ and $\lambda_{ei}=1.2\times 10^{-4} (T_{\rm e}^{\rm u})^2/n_{\rm e}^{\rm u}[10^{20}]$ from the connection length $L_{\parallel}$, the upstream electron density $n_{\rm e}^{\rm u}$ and the upstream electron temperature $T_{\rm e}^{\rm u}$ in a flux tube just outside the separatrix. Target parameters are also shown (index t). The connection length $L_{\parallel}$ is defined as the parallel distance between the outer midplane and outer target.}
\label{table1}
\end{center}
\end{table}

\subsection{Neutral species}

\begin{figure}[!h]
\centerline{\scalebox{0.35}{\includegraphics[clip]{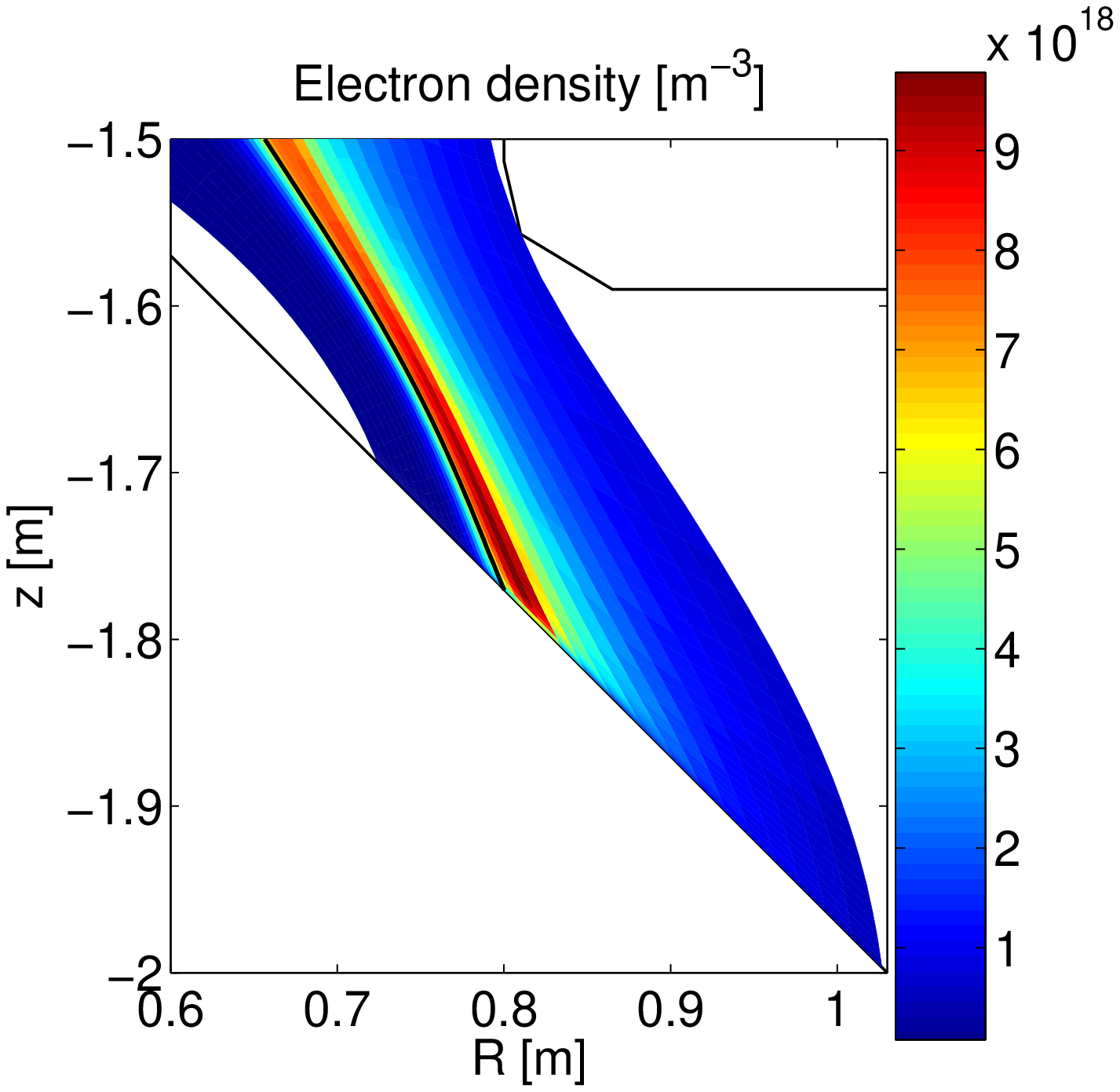}}
\scalebox{0.35}{\includegraphics[clip]{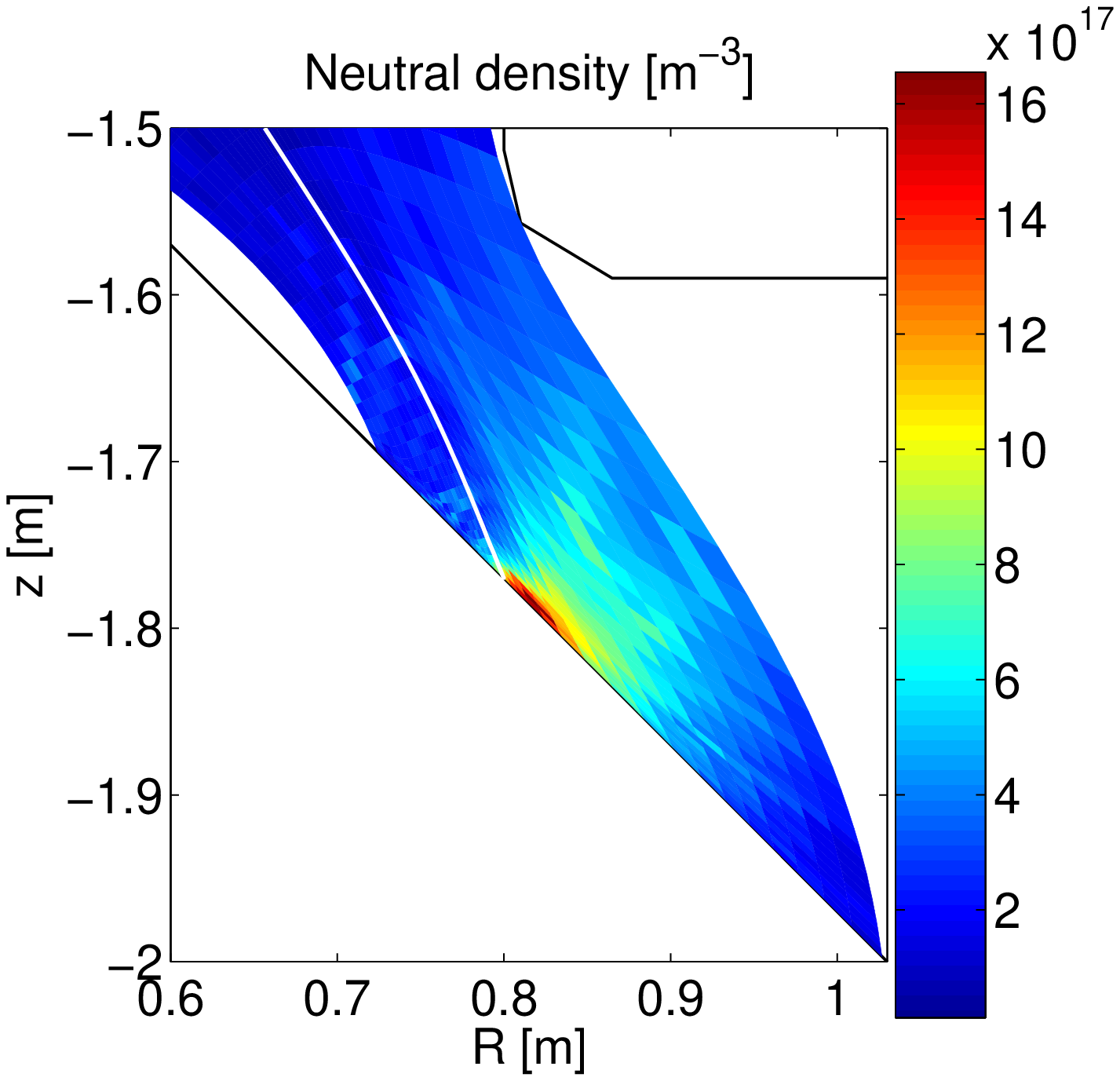}}
\scalebox{0.35}{\includegraphics[clip]{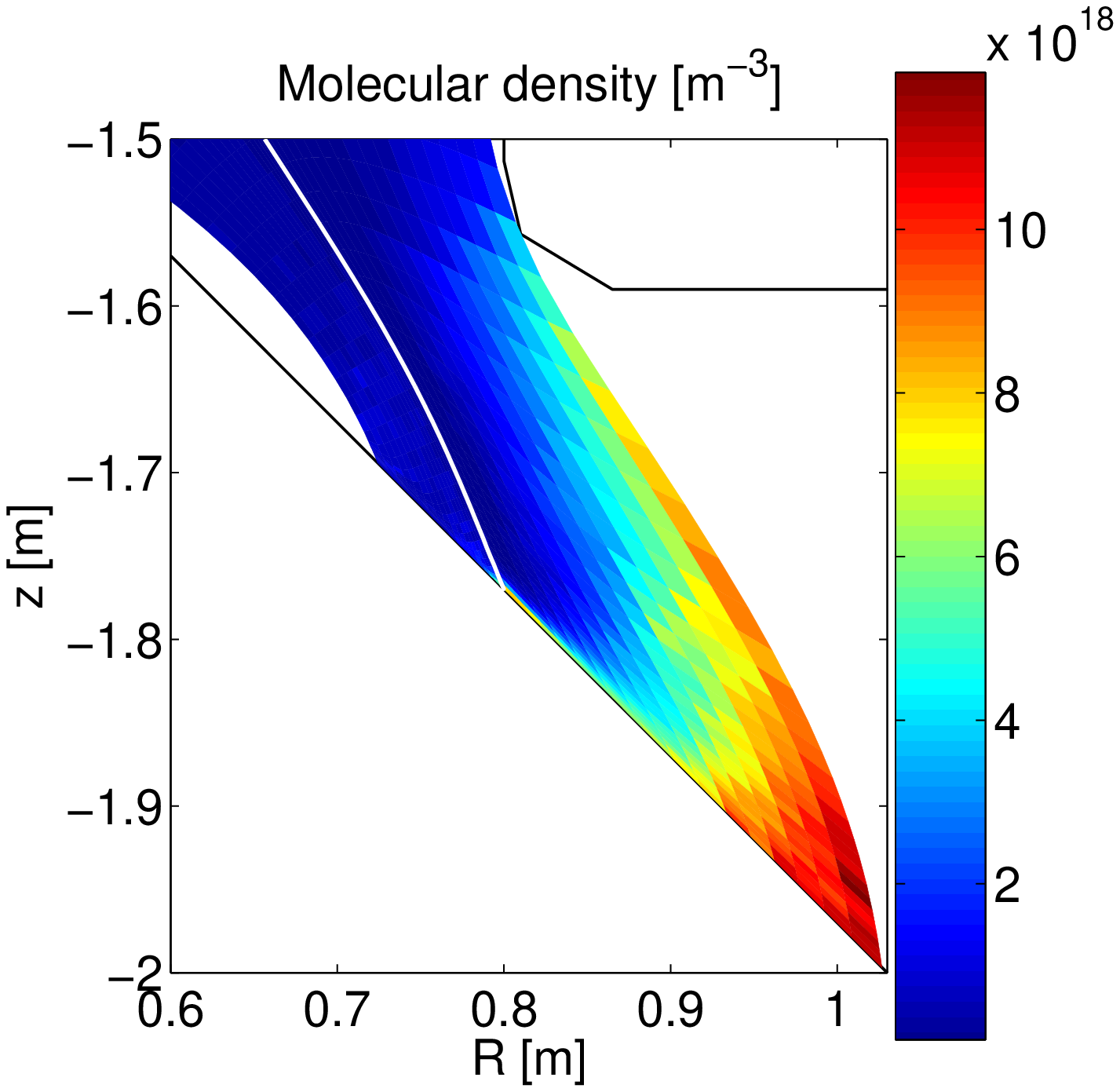}} 
} 
\caption{The CD configuration -- distribution of electrons, ${\rm D}_0$ and ${\rm D}_2$ in the divertor leg in the case (a).} \label{fig_distribution}
\end{figure}

\begin{figure}[!h]
\centerline{\scalebox{0.7}{\includegraphics[clip]{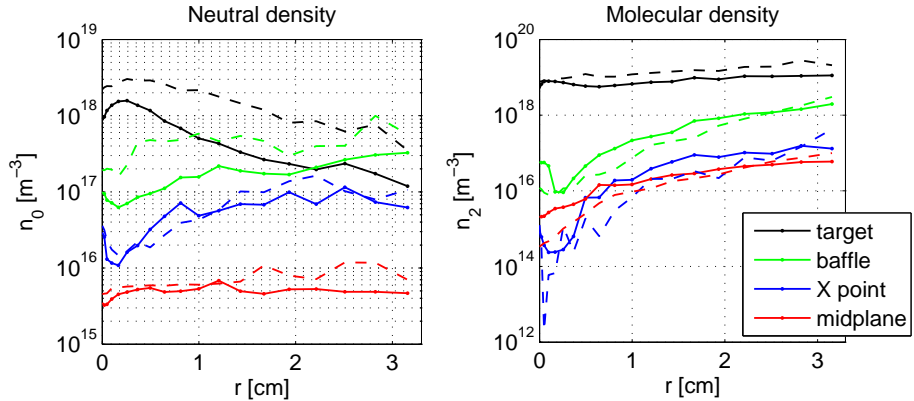}}} 
\caption{The CD configuration -- radial profiles (mapped to the midplane to have a common scale) of the ${\rm D}_0$ and ${\rm D}_2$ density for the case (a) (solid) and case (b) (dashed) at four poloidal locations in the outer SOL -- at the target (black), at the contact point of the plasma grid with the baffle (green), at the X-point (blue), at the midplane (red).} \label{fig_neutrals}
\end{figure}

The divertor closure with respect to neutral species or impurities is an important factor for the divertor performance. In the MAST-U design, a baffle structure has been introduced to separate the divertor region from the main plasma and to maintain neutral species below the X-point. The divertor closure is here evaluated by analyzing the distribution of neutrals and ionization sources. 
2D distribution of plasma species, neutrals and molecules in the divertor is displayed in Fig. \ref{fig_distribution} (for the meshed part of the vessel), illustrating regions where ionization and dissociation occur. In Fig. \ref{fig_neutrals}, radial profiles of neutral species are shown for both H-mode and L-mode cases (a) and (b) at four different poloidal locations in the outer SOL. Molecules and neutrals are fairly well separated from the main plasma as neutral densities at the baffle entrance are two orders of magnitude smaller than densities in front of the target plate. Important will be the effect of the divertor geometry on the divertor closure discussed in section \ref{sec_neut_sxd}, where also the difference between the MAST-U divertor with a baffle and the current MAST configuration with an open divertor is addressed. In simulations of the open MAST divertor, one typically finds similar neutral densities at the target as in Fig. \ref{fig_neutrals} (more neutrals escaping to the upper chamber, but less pumped in the divertor), but larger neutral densities at the midplane and in the core.

\subsection{Power losses and radiation}

As the divertor geometry influences the collisionality and the divertor closure with respect to neutrals and impurities, it also affects radiation losses. One seeks solutions to increase radiated power fractions as a mechanism to remove power from the plasma in the divertor volume before it is deposited at the target in a localized way.
In that context, the effect of the divertor geometry is investigated by analyzing the distribution of power losses in the divertor and the sensitivity of radiation to parameters such as the connection length.

\begin{figure}[!h]
\centerline{\scalebox{0.46}{\includegraphics[clip]{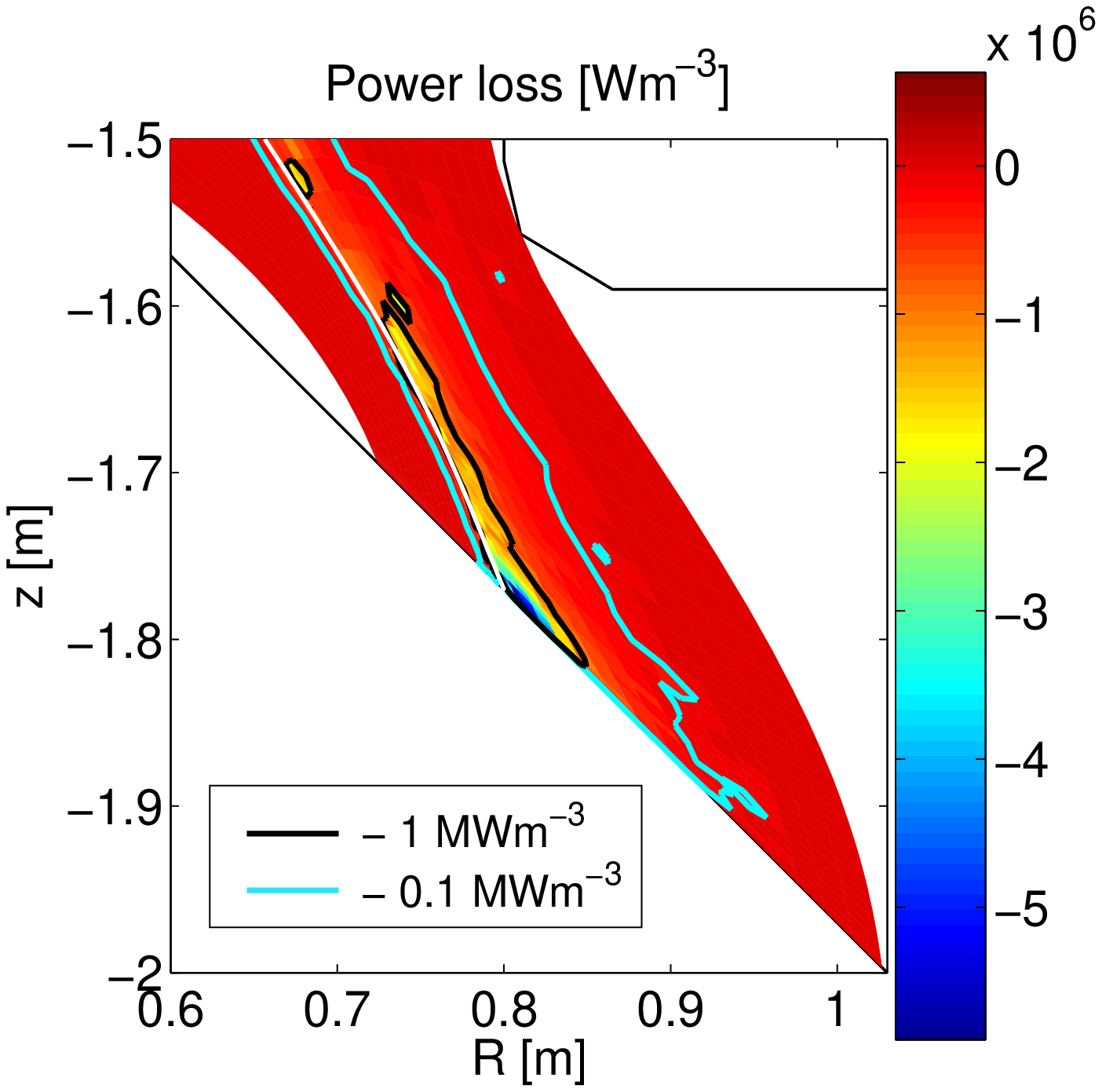}}
\scalebox{0.46}{\includegraphics[clip]{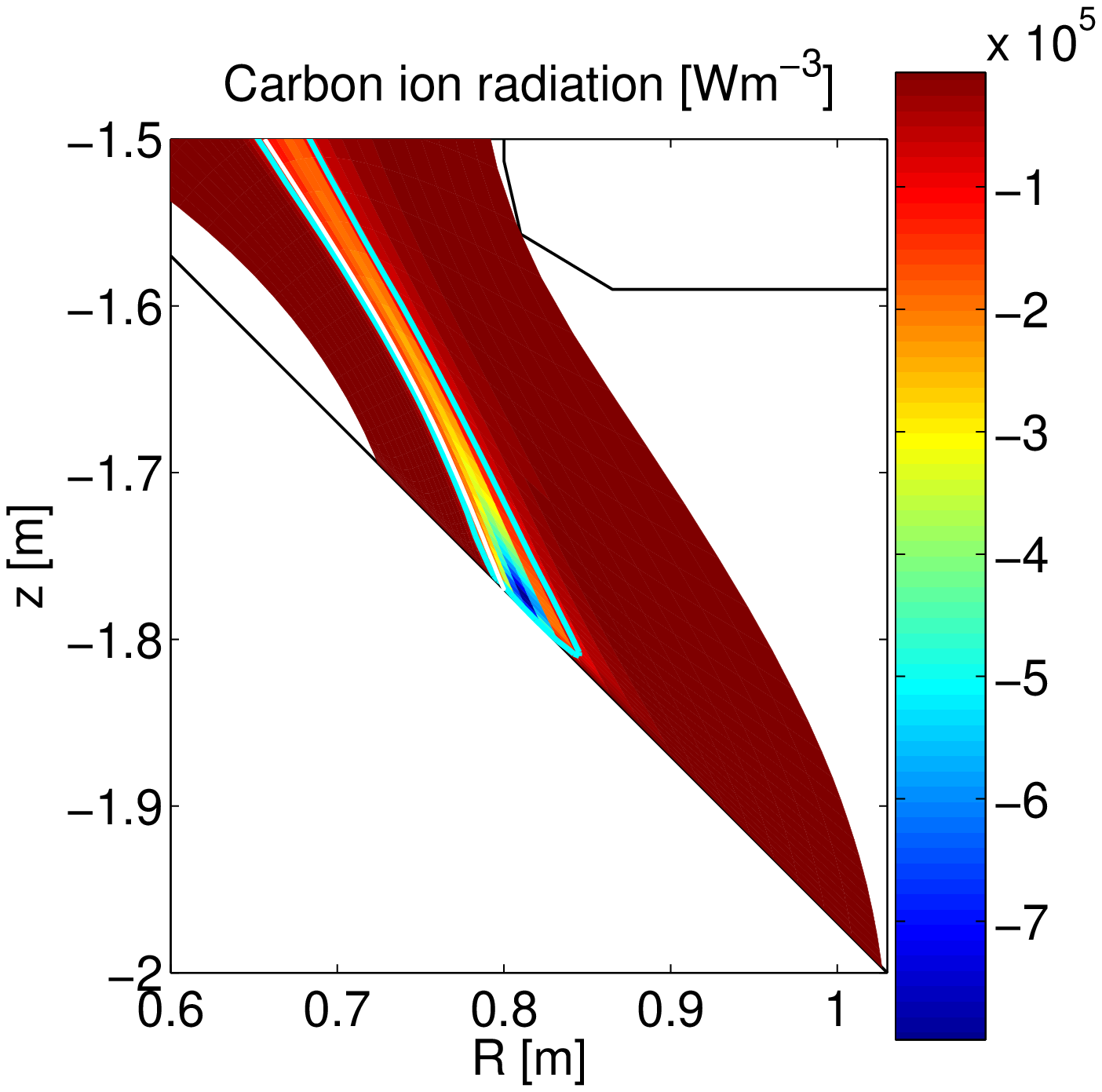}}
} 
\caption{The CD configuration -- (left) distribution of total power losses in the divertor for the case (a). The contour lines show regions in the divertor where the power loss exceeds a certain limit. (right) Distribution of carbon line radiation in the divertor.} \label{fig_volumetric}
\end{figure}

The volumetric power loss in the divertor leg of MAST-U is shown in Fig. \ref{fig_volumetric} left for the case (a), calculated as the total energy loss caused to the plasma due to impurity radiation and all plasma-neutral collisions including neutral radiation, charge-exchange and ionization processes (not possible to separate in the current version of SOLPS).
The total power loss is strongest in a small region around the strike point (see the black contour line that defines a region where the energy loss is larger than 1 MWm$^{-3}$), 
but a non-negligible power loss occurs in the divertor leg all the way up to the X-point (see the blue contour line that defines a region where the energy loss is larger than 0.1 MWm$^{-3}$). 90\% of the power loss in the divertor region displayed in Fig. \ref{fig_volumetric} takes place in the region defined by the blue contour. This region extends into the half SOL radially and up to the X-point poloidally and spreads over approximately 30\% of the meshed divertor volume. The temperature range is this region changes from one case to another, depending on other parameters such as electron and neutral densities. In Fig. \ref{fig_volumetric} on the right, carbon ion line radiation is shown separately, again extending up to the X-point. 

\begin{figure}[!h]
\centerline{\scalebox{0.5}{\includegraphics[clip]{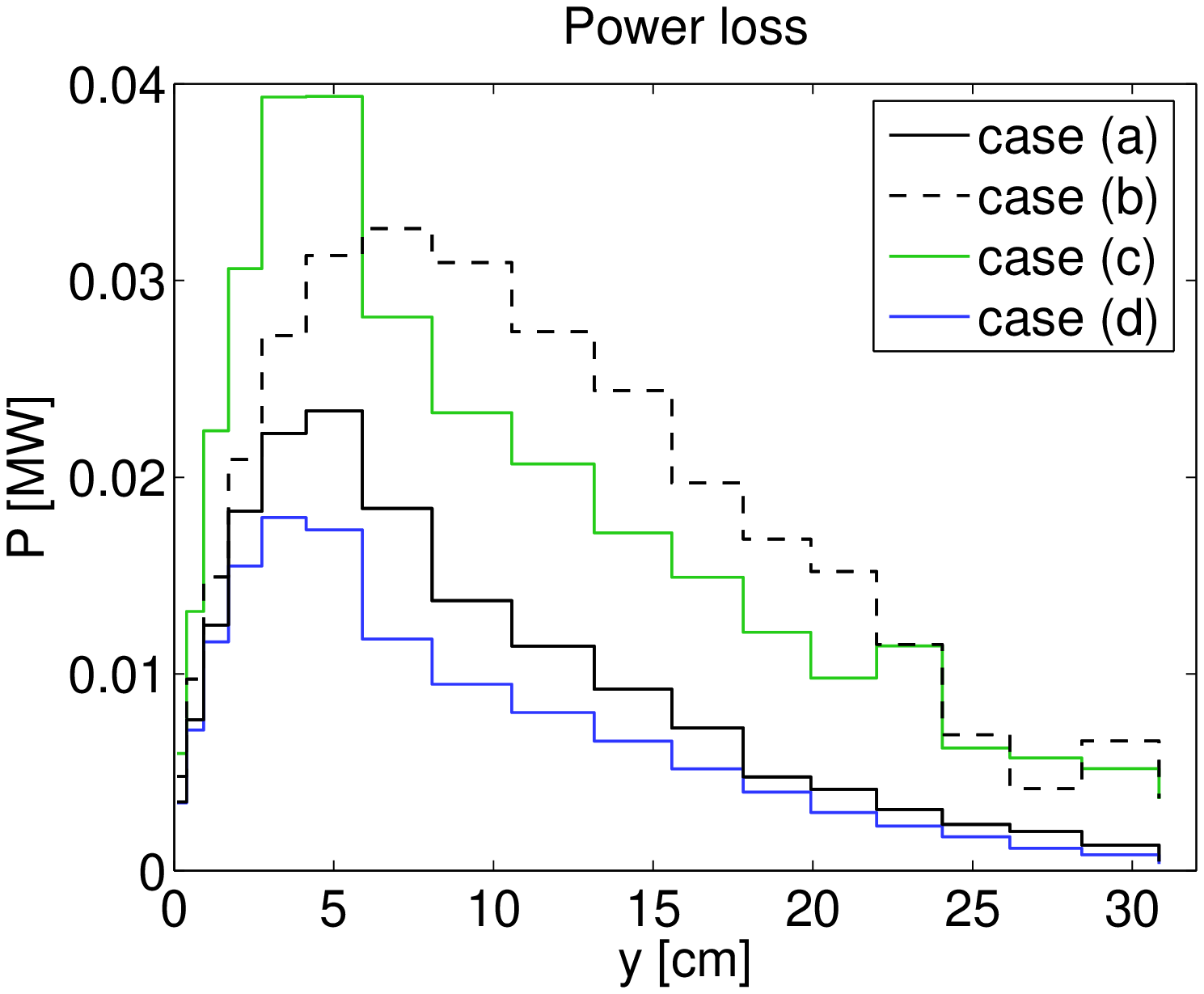}}
\scalebox{0.5}{\includegraphics[clip]{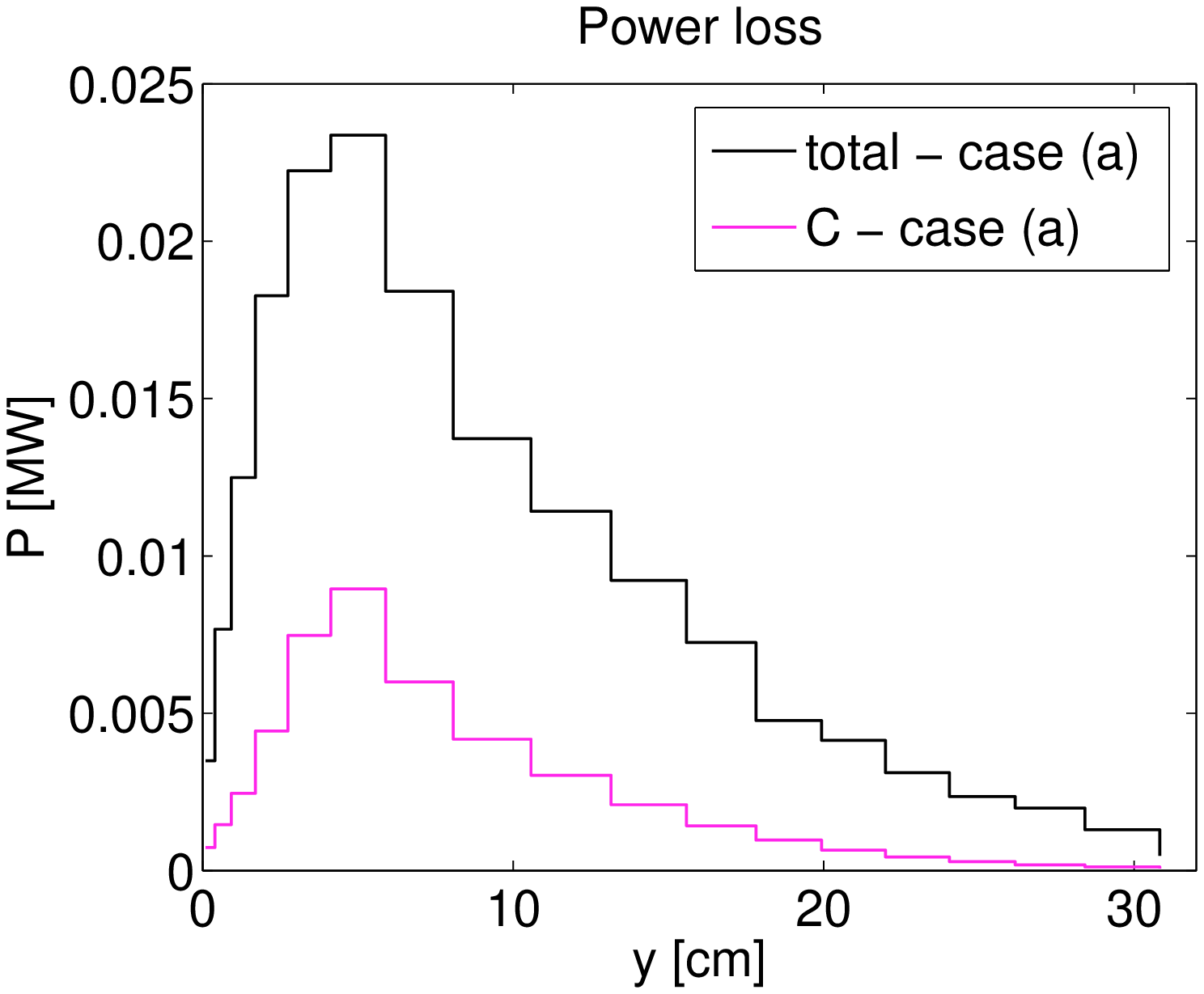}}
} 
\caption{The CD configuration -- (left) volumetric power losses in the outer SOL calculated for each flux tube individually. $y$ is the target coordinate with 0 corresponding to the separatrix location. All four cases are shown -- the H-mode case (a) (solid), the L-mode case (b) (dashed), the higher density case (c) (green) and the lower power case (d) (blue). (right) The total power loss and the loss due to carbon radiation are shown for the case (a).} \label{fig_volumetric2}
\end{figure}

For the L-mode case (b), the power loss region broadens radially up to 58\% of the meshed divertor volume. This is indicated in Fig. \ref{fig_volumetric2} on the left, where the power loss in the SOL is plotted for each flux tube separately as a function of the target coordinate. The cases (c) and (d) with increased density and reduced input power are shown as well. In the higher density case, due to increased collisionality, the power loss region broadens and the total power loss increases similarly to the case (b). Fig. \ref{fig_volumetric2} on the right separates carbon radiation and indicates that deuterium-based processes account for most of the power removal in the simulations at the given carbon sputtering yield. 

\begin{table}[!h]
\begin{center}
\begin{tabular}{|l|rrrr|rrrr|}
\hline
& \multicolumn{4}{|c|}{total} & \multicolumn{4}{|c|}{carbon} \\
    & (a) & (b) & (c) & (d) & (a) & (b) & (c) & (d) \\
\hline
$P_{\rm vol}$ in the core & 4.4\% & 4.4\% & 2.6\% & 4.0\% & 1.1\% & 1.9\% & 1.6\% & 1.7\%\\
$P_{\rm vol}$ in the SOL & 11.5\% & 21.7\% & 21.6\% & 18.2\% & 3.6\% & 7.1\% & 7.1\% & 5.5\%\\
$P_{\rm vol}$ in the outer divertors & 7.4\% & 15.3\% & 14.5\% & 11.2\% & 1.2\% & 3.3\% & 3.6\% & 2.2\%\\
$P_{\rm vol}$ in the blue region & 6.7\% & 14.9\% & 13.9\% & 9.8\% & 0.9\% & 2.8\% & 3.1\% & 1.7\%\\
$S$ of the blue region [m$^2$] & 0.043 & 0.081 & 0.066 & 0.037 & 0.016 & 0.031 & 0.021 & 0.013\\
$V$ of the blue region [m$^3$] & 0.209 & 0.403 & 0.329 & 0.177 & 0.076 & 0.145 & 0.099 & 0.059\\
\hline
\end{tabular}
\caption{The CD configuration -- the first four rows show the total volumetric power loss $P_{\rm vol}$ in terms of $P_{\rm inp}$ and the power loss due to carbon radiation separately in four SOLPS simulations (a), (b), (c) and (d) in different regions: (i) in the core (the closed field line part of the grid), (ii) in the SOL (the whole grid outside the separatrix), (iii) in the outer divertors (the region displayed in Fig. \ref{fig_volumetric}), (iv) in the blue region (defined in Fig. \ref{fig_volumetric} by the contour line). The bottom rows show the surface $S$ and the volume $V$ of this region. The total power loss is calculated from EIRENE as the total energy loss caused to the plasma due to all plasma-neutral interactions including neutral and impurity radiation, charge exchange, ionization.}
\label{table_volumetric}
\end{center}
\end{table}

Quantitatively, the power losses are presented in Tab. \ref{table_volumetric} 
separately for the core, the SOL, and the outer divertors. Tab. \ref{table_volumetric} shows that a large fraction of the power is lost in the region defined by the blue contour line in Fig. \ref{fig_volumetric}. The total volumetric power loss clearly grows in the L-mode case (b) and the higher density case (c). 
This is consistent with increased surface/volume from which the power loss occurs. 

To evaluate the effect of impurity radiation in the two different geometries, 
Tab. \ref{table_volumetric} separates the power loss caused by carbon ion line radiation. In CD, the power fraction radiated by carbon reaches 9\% in total (unseeded plasma with the chemical sputtering yield of 1\% for carbon), and will be compared with SXD in section \ref{sec_rad_sxd}. The effect of carbon concentration has also been tested by increasing the sputtering coefficient to 3\%. In CD, the power radiated by carbon in the SOL almost doubles from $4-7$\% to $6-13$\% and increases slightly in the core.

\subsection{Particle and energy fluxes}

\subsubsection{Target fluxes}
\label{sec_cd_tf}

\begin{figure}[!h]
\centerline{\scalebox{0.7}{\includegraphics[clip]{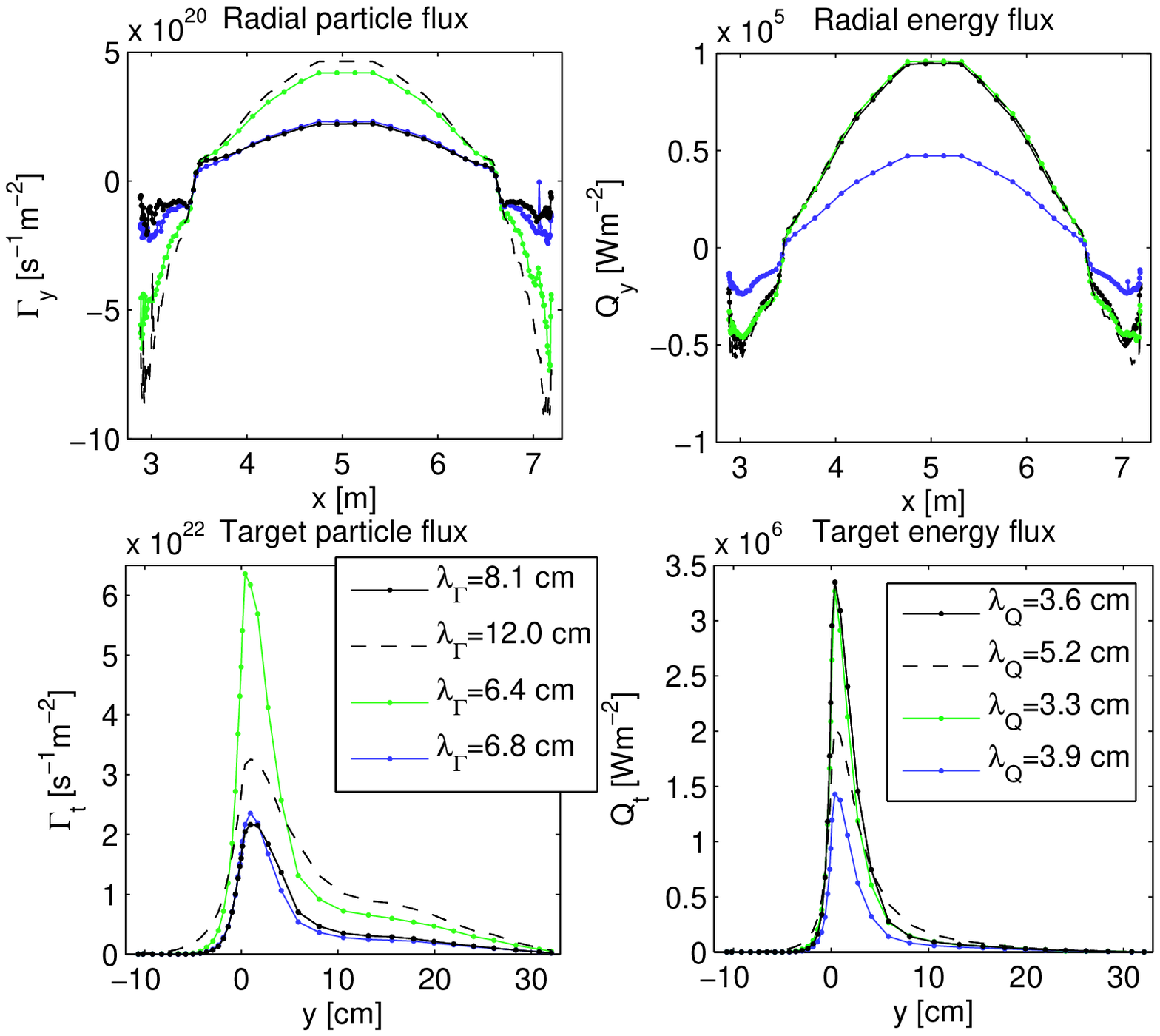}}
}
\caption{The CD configuration -- particle (deuterium ions) and power (deuterium ions plus electrons) loads at the outer target as functions of the target coordinate $y$ for the case (a) (solid), case (b) (dashed), case (c) (green) and case (d) (blue).} \label{fig_fluxes}
\end{figure}

Particle and power loads to the outer target are shown in Fig. \ref{fig_fluxes} for all four simulations. The target wetted area is characterized by the profile width at the target defined as 
\begin{equation}
\lambda_{\Gamma}^{\rm t}=\frac{\int \Gamma_{\rm t} {\rm d}y}{\Gamma_{\rm t}^{\rm max}}, \quad \lambda_Q^{\rm t}=\frac{\int Q_{\rm t} {\rm d}y}{Q_{\rm t}^{\rm max}}.
\end{equation}
$\lambda_{\Gamma}^{\rm t}\approx 8.1$ cm and $\lambda_{Q}^{\rm t}\approx 3.6$ cm are found for the case (a) (index t as target) and a factor of 1.5 larger for the case (b). 
Mapped to the midplane as $\lambda^{\rm u}=\lambda^{\rm t}/\langle {\rm d}y/{\rm d}r \rangle$, the power decay length in the H-mode case (a) is $\lambda_{Q}^{\rm u}\approx 0.3$ cm and the particle flux width is $\lambda_{\Gamma}^{\rm u}\approx 0.6$ cm. To calculate the flux expansion factor $\langle {\rm d}y/{\rm d}r \rangle$, averaging over $\lambda_{Q}^{\rm t}\approx 4$ cm is used for the energy flux and averaging over $\lambda_{\Gamma}^{\rm t}\approx 8$ cm is used for the particle flux.

The profile width broadens in the case (b) for both particles and energy due to increased radial diffusivities, consistently with reduced peak power load $Q_{\rm t}$. The higher density case (c) has no effect on the power load (in attached conditions), while it causes larger particle load $\Gamma_{\rm t}$. For two times lower input power in the case (d), $Q_{\rm t}$ is reduced by a factor of 2, while $\Gamma_{\rm t}$ remains the same.

\subsubsection{Flux expansion}

In novel divertor configurations such as SXD, an expansion of the plasma in the divertor region by magnetic geometry is used as one of the mechanisms to reduce the target energy flux. 
The effective flux expansion $\langle FX\rangle$ between the midplane and target locations for the CD case is shown in Tab. \ref{table_fx} and will be compared with SXD in section \ref{sec_fx_sxd}. The total flux expansion of 6.0 is separated into the toroidal flux expansion $\langle FX_{\rm tor}\rangle$ of 0.6 (the outer strike point at smaller $R$ than the outer midplane SOL), the poloidal flux expansion $\langle FX_{\rm pol}\rangle$ of 3.4 and the broadening caused by the target tilting $\langle FX_{\rm tilt}\rangle$ of 2.9 (defined in Appendix).
The flux expansion factors are calculated using averaging over 4 cm which corresponds to $\lambda_{Q}^{\rm t}$ in the H-mode case (a) and they change only little for averaging over 8 cm (corresponding to $\lambda_{\Gamma}^{\rm t}$) which is also shown in Tab. \ref{table_fx}. 

Compared to the current MAST configuration with a horizontal target where the effective flux expansion is $\langle FX\rangle \approx 2.3$, the major change is in $\langle FX_{\rm tilt}\rangle$ which increases from approximately 1 to 3, while $\langle FX_{\rm tor}\rangle$ and $\langle FX_{\rm pol}\rangle$ do not change much. Thanks to the target tilting in CD, the peak $\Gamma_{\rm t}$ and $Q_{\rm t}$ are reduced compared to the current MAST divertor for the same total particle flux and power arriving to the target.

\begin{table}[!h]
\begin{center}
\begin{tabular}{lrrrrrr}
 & $\langle FX\rangle$ & $\langle FX_{\rm tor}\rangle$ & $\langle FX_{\rm pol}\rangle$ & $\langle FX_{\rm tilt}\rangle$ & $\langle {\rm d}y/{\rm d}r\rangle$ \\
\hline
average over $\lambda_{Q}$ & 6.02 & 0.61 & 3.40 & 2.89 & 11.61 \\
average over $\lambda_{\Gamma}$ & 6.68 & 0.62 & 3.39 & 3.16 & 12.84 \\
\hline
\end{tabular}
\caption{
The CD configuration -- flux expansion factors between the outer midplane and outer target.
}
\label{table_fx}
\end{center}
\end{table}

\subsubsection{Global balance}

Global power balance and particle fluxes are calculated in Tab. \ref{table2}. Most of the power and particles cross the separatrix at the outer side (note that poloidally uniform $D_{\perp}$ and $\chi_{\perp}$ are assumed for simplicity in the model) and are therefore deposited mainly at the outer targets (connected double null). In the H-mode case (a) with low $D_{\perp}$ and $\chi_{\perp}$ and a narrow SOL, 76\% of $P_{\rm inp}$ is deposited at the outer targets. In the cases (b,c,d), it is  less than 70\%, while more power goes to the outer wall or is radiated in the SOL. In all cases, the outer targets receive most of the power ($60-80$\%) and the energy fluxes to the wall are only $1-2$\% of $P_{\rm inp}$.
The power balance is similar in simulations of the current MAST divertor in terms of the power deposited at the targets and the total power loss, with the difference that there is more power radiated in the core and less in the SOL in MAST with an open divertor compared to MAST-U with CD.

\begin{table}[!h]
\begin{center}
\begin{tabular}{lcccc}
    & (a) & (b) & (c) & (d) \\
\hline
$P_{\rm sol,in}$, $P_{\rm sol,out}$ &  11.9\%, 82.7\% & 12.4\%, 83.4\% & 12.8\%, 84.0\% & 12.3\%, 82.6\% \\
$P_{\rm t,in}$, $P_{\rm t,out}$ &  13.3\%, 75.9\% & 12.0\%, 65.3\% & 12.0\%, 67.9\% & 11.8\%, 68.7\% \\
$P_{\rm wall}$, $P_{\rm pfr}$ &  1.1\%, 0\%  & 2.2\%, 0\% & 2.0\%, 0\% & 1.8\%, 0\% \\
$P_{\rm vol}$ &  16.4\% & 24.2\% & 22.1\% & 23.0\% \\
$P_{\rm vis}$ &  6.7\% & 3.7\% & 4.0\% & 5.3\% \\
\hline
$F_{\rm sol}$ & $4.3\times 10^{21}$ s$^{-1}$ & $8.4\times 10^{21}$ s$^{-1}$ & $7.9\times 10^{21}$ s$^{-1}$ & $4.4\times 10^{21}$ s$^{-1}$\\
$F_{\rm sol,out}$ & 81.0\% & 84.9\% & 82.3\% & 81.4\%  \\
$F_{\rm t,out}$ & $2.0\times 10^{22}$ s$^{-1}$ & $4.3\times 10^{22}$ s$^{-1}$ & $4.4\times 10^{22}$ s$^{-1}$ & $1.8\times 10^{22}$ s$^{-1}$  \\
$F_{\rm wall,out}$ & 1.9\% & 2.6\% & 2.0\% & 2.1\% \\
\hline
\end{tabular}
\caption{The CD configuration -- the upper part of the table shows the power balance in the simulations. Expressed in \% of $P_{\rm inp}$, the table shows: 
$P_{\rm sol,in}$ and $P_{\rm sol,out}$ -- the power crossing the separatrix on the inner or outer side, $P_{\rm t,in}$ and $P_{\rm t,out}$ -- the power deposited at the inner and outer targets, $P_{\rm wall}$ -- the power deposited at the outer boundary of the SOLPS domain (both inboard and outboard wall), $P_{\rm pfr}$ -- the power crossing the private flux region boundaries, $P_{\rm vol}$ -- the volumetric power loss, $P_{\rm vis}$ -- the viscous heating. It holds $P_{\rm inp}=P_{\rm vol}+P_{\rm t,in}+P_{\rm t,out}+P_{\rm wall}+P_{\rm pfr}-P_{\rm vis}$. 
The lower part of the table shows the particle flux crossing the separatrix $F_{\rm sol}$ from which $F_{\rm sol, out}$ crosses the separatrix at the outer side. $F_{\rm t,out}$ is the flux deposited at the outer targets and $F_{\rm wall, out}$ is the flux deposited at the outer wall expressed in terms of  $F_{\rm t,out}$. All fluxes are calculated as sums over all ion species. Note that here the volumetric power loss $P_{\rm vol}$ is calculated from the balance between the input power and the fluxes deposited at the solid surfaces, see the equality above. This calculation of $P_{\rm vol}$ leads to a similar result as in Tab. \ref{table_volumetric}, proving the energy balance in the simulation is satisfied.}
\label{table2}
\end{center}
\end{table}

The applied version of SOLPS does not enable to separate the total power loss into individual components caused by different collision processes. Separately can be calculated impurity radiation and Bremsstrahlung. Changes have been made to the code to separate also neutral radiation and results are presented in \cite{Eva3}. Apart from radiation, it is assumed that the dominant power loss process are charge exchange and ionization. Bremsstrahlung is negligible in MAST (0.1\% here). The power fraction radiated by carbon is $5-9$\% of $P_{\rm inp}$ (accounts for approximately $30-40$\% of the total power loss) and increases by a factor of 2 with increased sputtering yield (3\%) to $8-16$\% (corresponding to $40-45$\% of the total power loss). This increases the total power losses and reduces the power to the outer targets. The combined effect of the sputtering yield and SXD will be discussed in section \ref{sec_gb_sxd}.

In all simulations, the divertor baffle receives only a small fraction of the particle flux with respect to the flux deposited at the target and therefore most of the recycling and sputtering takes place in the divertor. On the other hand, the divertor remains closed for neutrals thanks to the closed design with a baffle (see later in Tab. \ref{table_atomfluxes}).

\section{Conventional versus Super-X divertor performance}

\subsection{Super-X geometry}

\begin{figure}[!h]
\centering
\begin{tabular}{ll}
\multirow{2}{*}[2.5cm]{
\scalebox{0.5}{\includegraphics[clip]{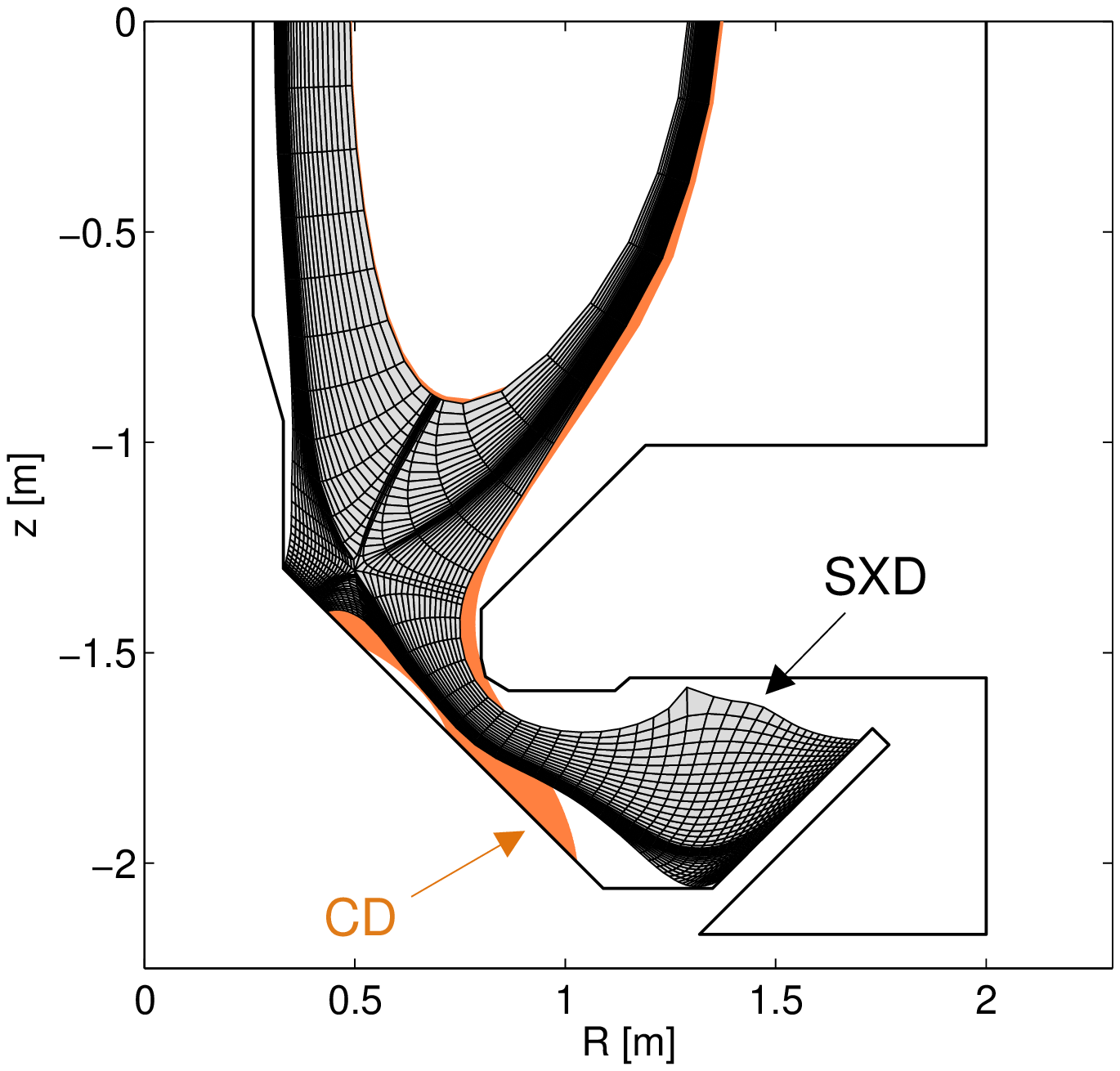}}
} & 
\scalebox{0.4}{\includegraphics[clip]{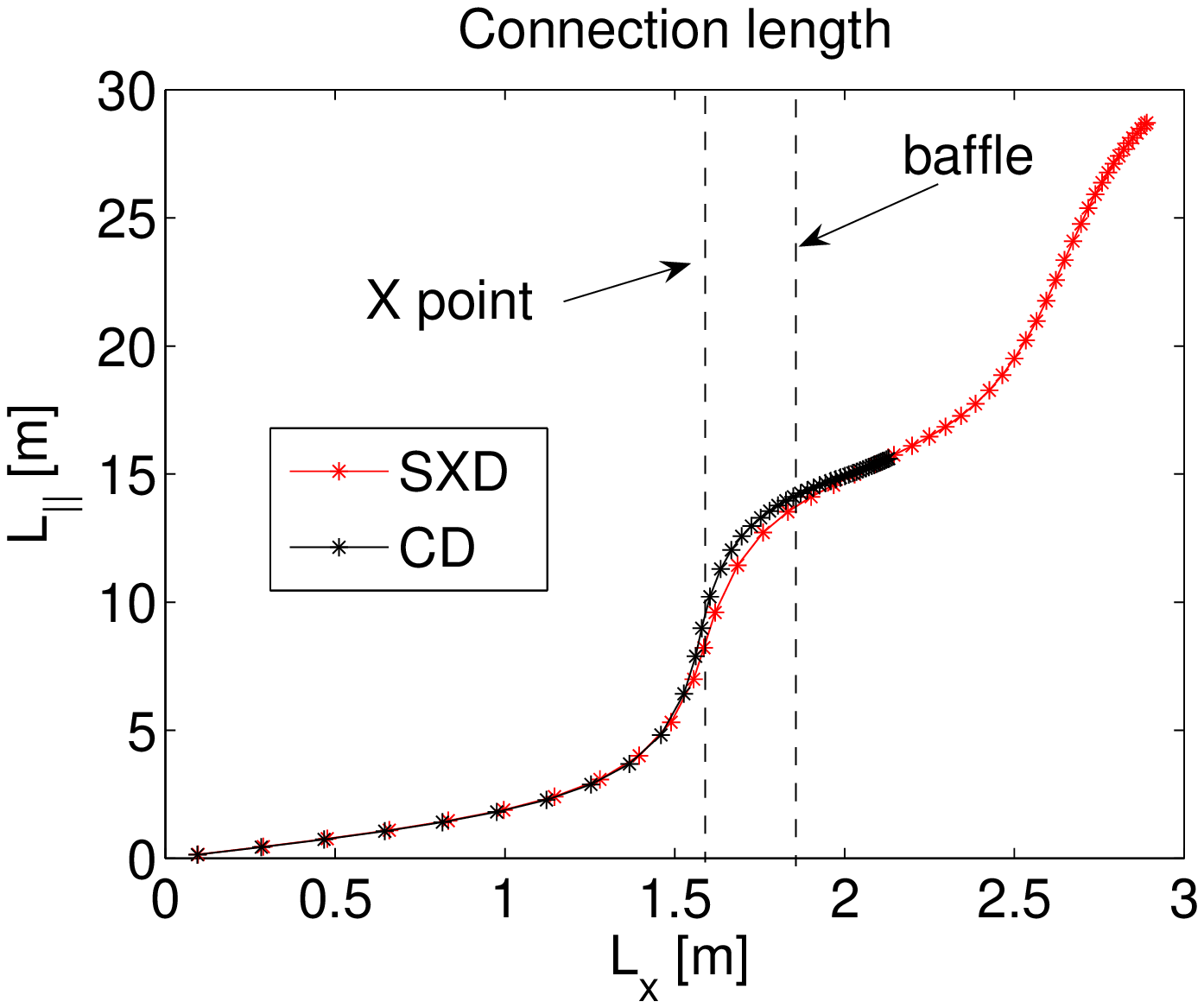}} \\
& \scalebox{0.4}{\includegraphics[clip]{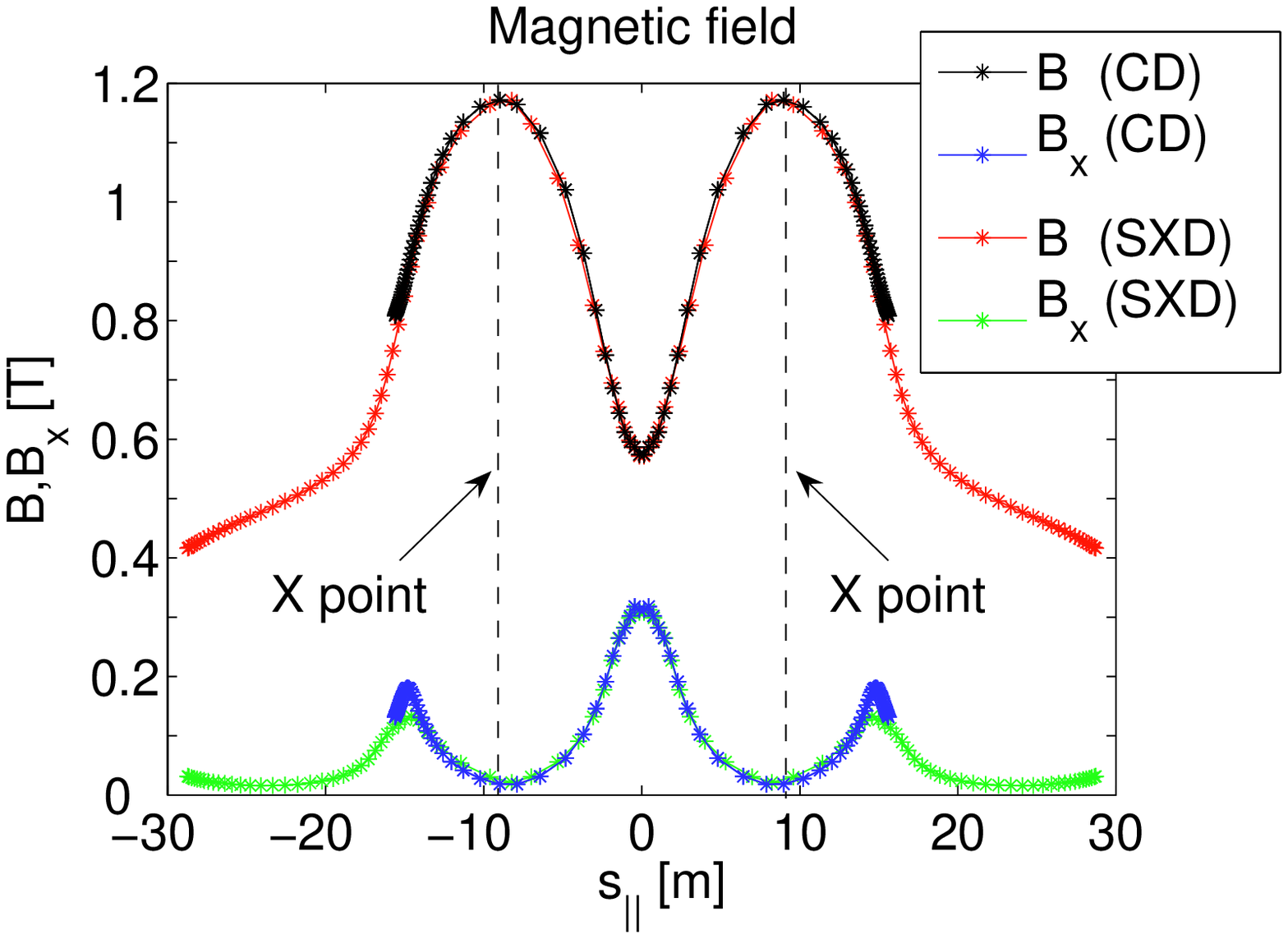}}
\end{tabular}
\caption{On the left, a SOLPS grid for the SXD geometry compared against the CD geometry. On the right, (top) the connection length $L_{\parallel}$ as a function of the poloidal distance from the outer midplane to the outer target $L_x$, (bottom) the total magnetic field $B$ and the poloidal magnetic field $B_x$ compared in the two configurations. $B$ and $L_{\parallel}$ are shown on a flux tube in the near SOL which is defined as the flux tube with maximum power load at the target.} \label{fig_grid_sxd}
\end{figure}

One of the main design features of the new divertor in MAST-U are additional poloidal coils in the divertor region (see e.g. \cite{Lisgo}) that allow to vary the connection length in the divertor, the strike point location along the target and the expansion of the plasma in the divertor. This enables the Super-X magnetic geometry. One of the considered connected double null configurations is shown in Fig. \ref{fig_grid_sxd}. The double null topology benefits from the separation of the inner and outer SOL -- most of the power goes to the outer side, where the divertor is modified to distribute the power over larger target area. In Fig. \ref{fig_grid_sxd}, the magnetic configuration on the inner side does not change and in the simulation only a small change is found in plasma parameters on the inner side, therefore we focus mainly on results in the outer SOL. 

The increased target wetted area in SXD is achieved by magnetic flux expansion, see section \ref{sec_fx_sxd} and Appendix, which is a combination of poloidal flux expansion (reduced poloidal magnetic field in the vicinity of poloidal coils) and toroidal flux expansion (reduced toroidal magnetic field at larger radius). The corresponding drop of 
the total magnetic field $B$ and its poloidal component $B_x$  is shown in Fig. \ref{fig_grid_sxd} for a flux tube in the near outer SOL.

Apart from the reduction of $Q_{\rm t}$ through magnetic geometry, larger volumetric power losses in SXD are expected due to higher collisionality and better divertor closure (higher neutral pressure in the divertor and reduced temperatures, i.e. stronger radiation rate), see sections \ref{sec_neut_sxd} and \ref{sec_rad_sxd}.

The connection length increases from 20 to 35 m in a flux tube just outside the separatrix (at $r\approx 0.1$ mm), giving an opportunity for more power to be removed along the field line. The major increase occurs around the X-points and in the extended divertor region where the poloidal magnetic field is weak. It is expected that longer $L_{\parallel}$ leads to a broadening of the power deposition profile, hence a reduction of the peak load, as particles and energy have longer distance to travel before reaching the target, while they traverse radially (experimentally observed in \cite{Petrie}). This effect will be discussed in section \ref{sec_lpar_effect}.

\subsection{Plasma parameters}
\label{sec_par_sxd}

Four SOLPS simulations with parameters listed in Tab. \ref{table_runs} are repeated for the SXD configuration and compared with those in section \ref{sec_cd} for CD. 
The same radial transport coefficients are used (see Fig. \ref{fig_dchi}), as we do not yet have a basis for a different radial transport. We will see that such assumption gives similar conditions in the core and upstream SOL and this will represent the starting point of our comparison. At the same time, this approach gives an opportunity to separate changes in the divertor region from discrepancies in the upstream SOL.

\begin{figure}[!h]
\centerline{\scalebox{0.68}{\includegraphics[clip]{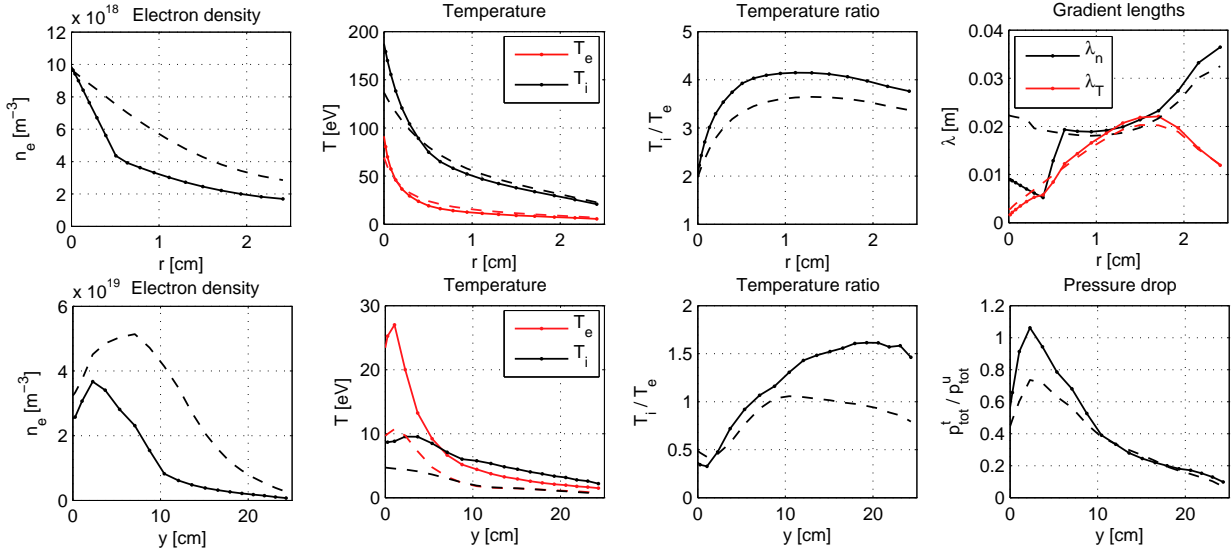}}
}
\caption{The SXD configuration -- (top) midplane radial profiles of (i) the electron density, (ii) the electron and ion temperature, (iii) the ratio of the temperatures, (iv) the gradient lengths of the density and temperature. (bottom) Target radial profiles of (i) the electron density, (ii) the electron and ion temperature, (iii) the ratio of the temperatures, (iv) the pressure drop along the SOL measured as the ratio of the target and upstream total pressure $p_{\rm tot}=m_{\rm i}n_{\rm e}u_{\parallel}^2+n_{\rm e}k(T_{\rm e}+T_{\rm i})$. The SXD case (a) is shown as a solid line, while the case (b) is shown as a dashed line. Compare with Fig. \ref{fig_radial}.} \label{fig_radial_sxd}
\end{figure}

\begin{figure}[!h]
\centerline{\scalebox{0.68}{\includegraphics[clip]{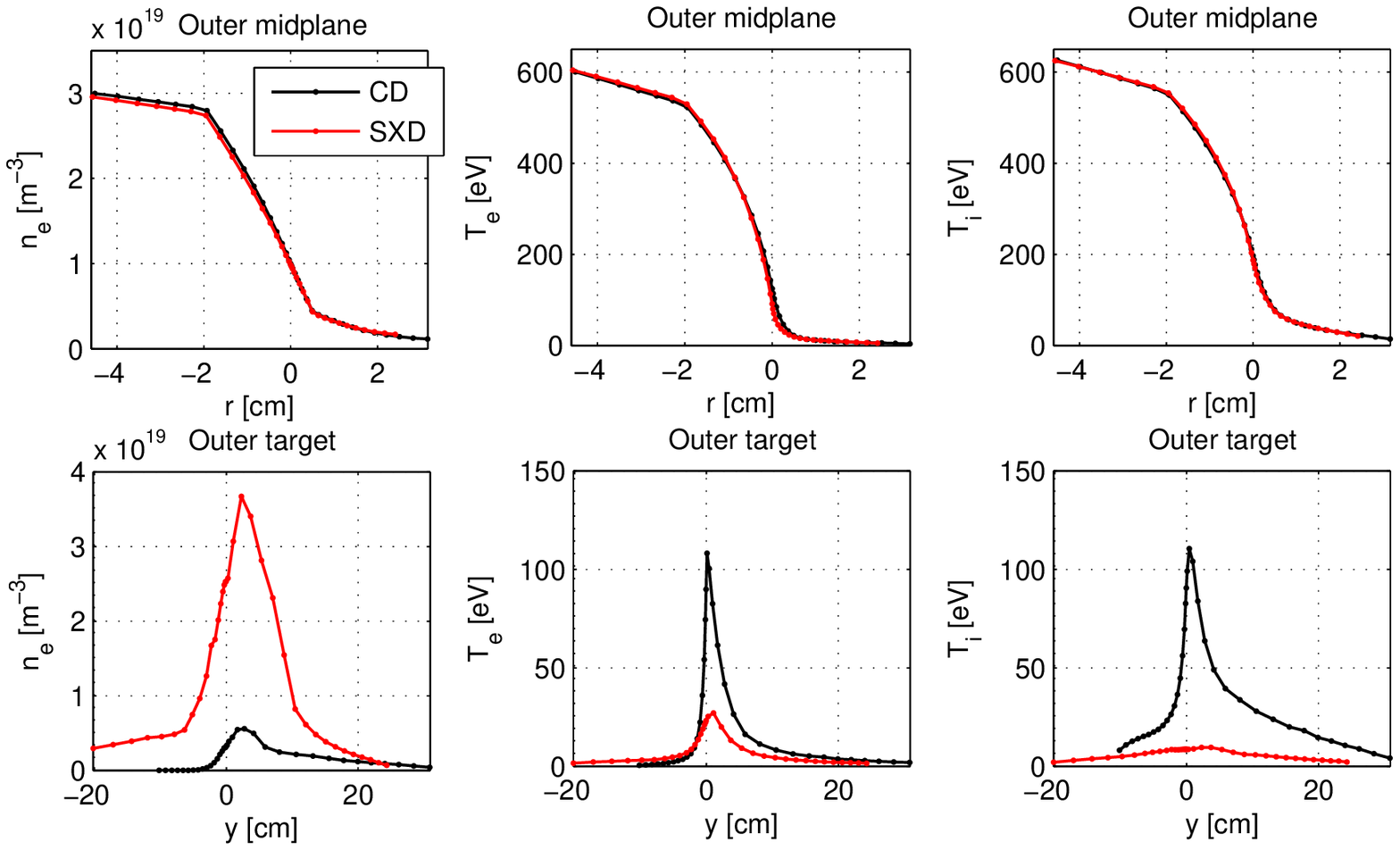}}
}
\caption{Direct comparison of radial profiles between CD (black) and SXD (red) for the case (a) in the outer SOL.} \label{fig_radial_comp_outer}
\end{figure}

Radial profiles of plasma parameters for the H-mode and L-mode cases (a) and (b) are shown in Fig. \ref{fig_radial_sxd}.
While the midplane densities and temperatures are comparable to those in Fig. \ref{fig_radial} for CD, a strong reduction of the temperatures in the divertor occurs in SXD as the result of collisional cooling and flux expansion, accompanied by an increase of the target density. The temperature ratio at the outboard midplane of SXD goes up to $T_{\rm i}/T_{\rm e}\approx 4$ (similar to CD), while it is reduced at the target compared to the CD case (stronger thermal coupling at larger $n$ and smaller $T$ as the result of the energy exchange between electrons and ions described by $Q_{\rm ei}\propto n^2(T_{\rm e}-T_{\rm i})/T_{\rm e}^{3/2}$). 
The SOL width at the outboard midplane is comparable between CD and SXD for the same radial transport coefficients as shown by the radial decay lengths in Figs. \ref{fig_radial} and \ref{fig_radial_sxd}, and also later by a direct comparison of the outer midplane profiles in Fig. \ref{fig_radial_comp_outer}. In the case (a) with reduced radial transport, the plasma is attached, while in the case (b), a pressure drop along the field lines starts to form due to momentum losses as we approach lower target temperatures of $T_{\rm e}\approx 10$ eV. Tab. \ref{table1_sxd} 
gives a summary of plasma parameters in SXD in all four simulations, which range from attached plasma in the case (a) to cases where the outer leg is not far from the detachment limit of $T_{\rm e}\approx 5$ eV (b,c,d). 

\begin{table}[!h]
\begin{center}
\begin{tabular}{lrrrrrrrrr}
    & $L_{\parallel}$ & $n_{\rm e}^{\rm u}$ & $T_{\rm e}^{\rm u}$ & $T_{\rm i}^{\rm u}$ & $n_{\rm e}^{\rm t}$ & $T_{\rm e}^{\rm t}$ & $T_{\rm i}^{\rm t}$ & $\lambda_{\rm ei}$ & $\nu_{\rm e}$ \\   
    & [m] & [10$^{19}$ m$^{-3}$] & [eV] & [eV] & [10$^{19}$ m$^{-3}$] &  [eV] &  [eV] & [m] &\\  
\hline
(a) & 35 & 1.0 & 81  & 179 & 3.7 & 27.0 & 9.5 & 8.1 & 4.4 \\
(b) & 35 & 1.0 & 65 & 134 & 5.1 & 10.7 & 5.0 & 5.2 & 6.8 \\
(c) & 35 & 1.5 & 71 & 138 & 8.9 & 13.5 & 6.3 & 3.9 & 9.0 \\
(d) & 35 & 0.9 & 58 & 118 & 4.0 & 6.8 & 4.3 & 4.5 & 7.8 \\
\hline
\end{tabular}
\caption{The SXD configuration -- the collisionality of the SOL plasma $\nu_{\rm e}$ and the mean free path $\lambda_{\rm ei}$ calculated as $\nu_{\rm e}=L_{||}/\lambda_{ei}$ and $\lambda_{ei}=1.2\times 10^{-4} (T_{\rm e}^{\rm u})^2/n_{\rm e}^{\rm u}[10^{20}]$ from the connection length $L_{\parallel}$, the upstream electron density $n_{\rm e}^{\rm u}$ and the upstream electron temperature $T_{\rm e}^{\rm u}$ in a flux tube just outside the separatrix. Target parameters are also shown (index t). Compare with Tab. \ref{table1}.}
\label{table1_sxd}
\end{center}
\end{table}

\begin{figure}[!h]
\centerline{\scalebox{0.68}{\includegraphics[clip]{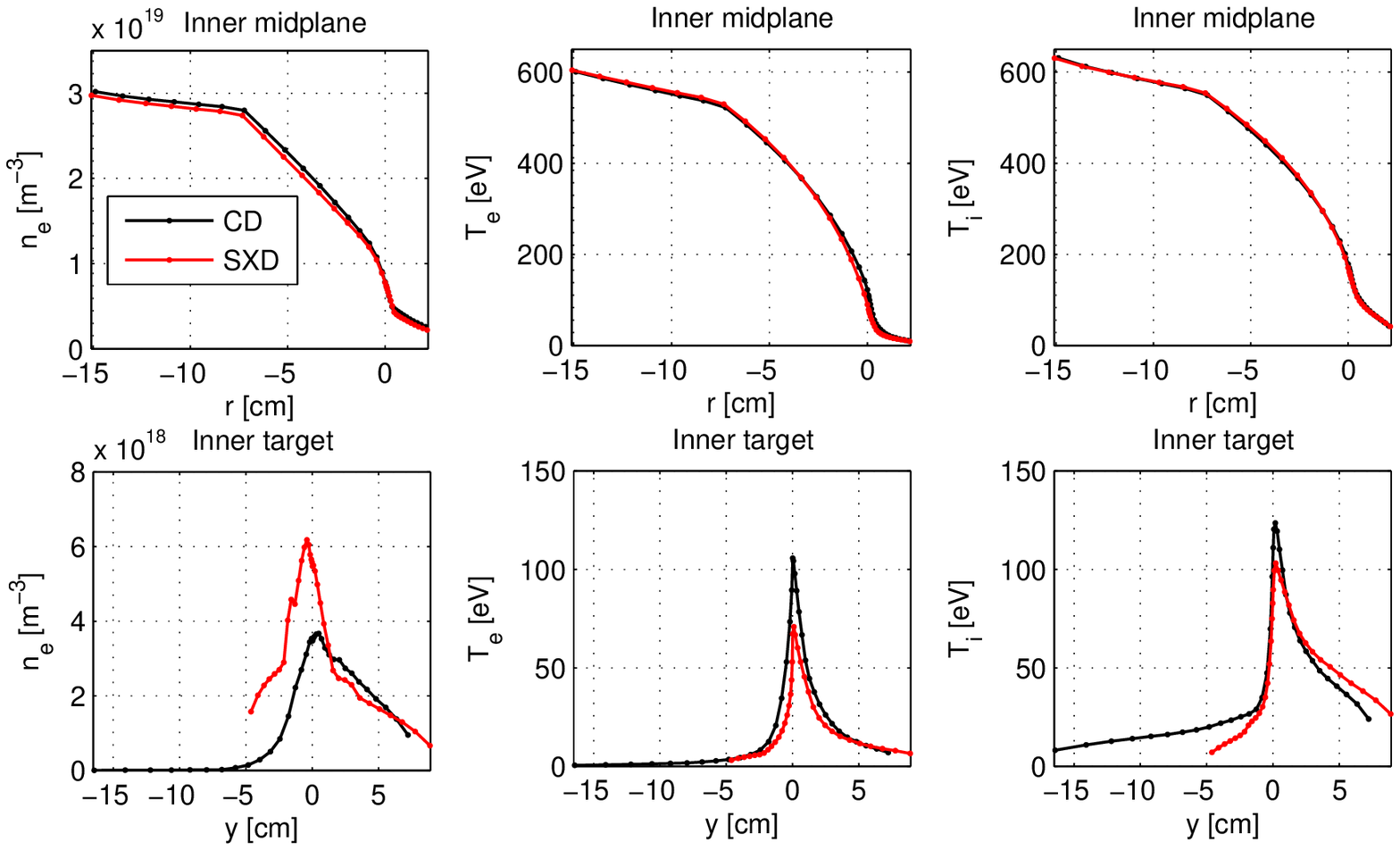}}
}
\caption{Direct comparison of radial profiles between CD (black) and SXD (red) for the case (a) in the inner SOL.} \label{fig_radial_comp_inner}
\end{figure}

A direct comparison of the radial profiles for the case (a) is presented in Figs. \ref{fig_radial_comp_inner} (inner SOL) and \ref{fig_radial_comp_outer} (outer SOL). Plasma parameters are strongly modified in the extended outer divertor which stays isolated from the rest of the plasma where the densities and temperatures remain rather unchanged. A small change in the inner divertor is caused by larger particle flux from the outer divertor across the private flux region resulting from larger density in the outer leg in SXD. 

\begin{figure}[!h]
\centerline{\scalebox{0.68}{\includegraphics[clip]{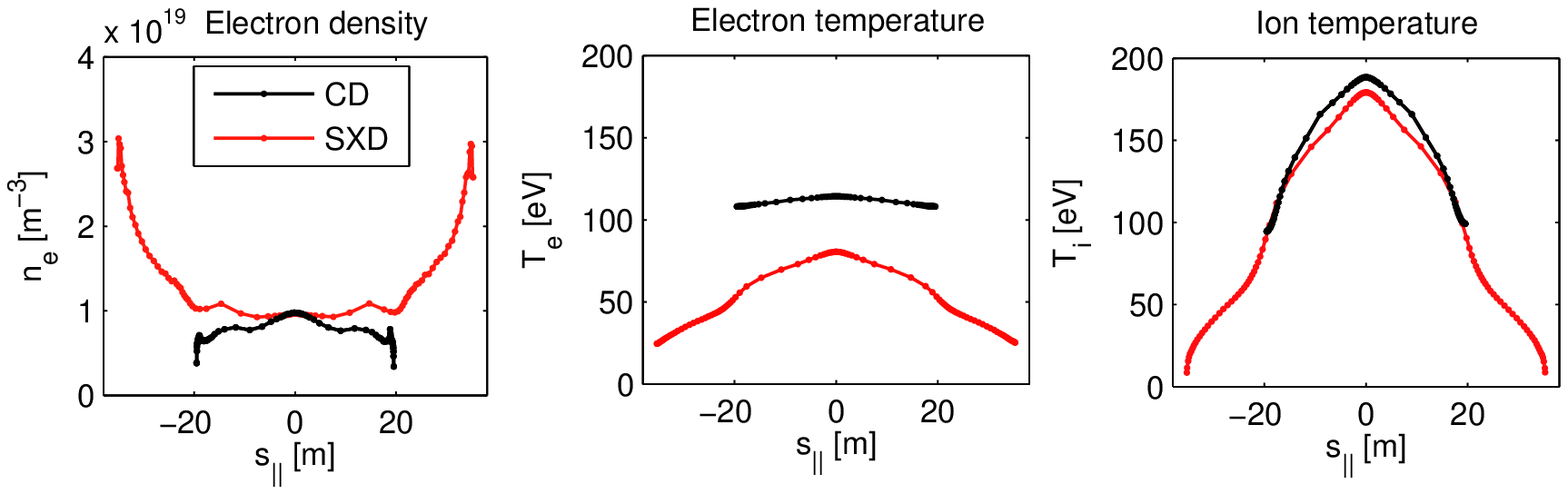}}
}
\caption{Direct comparison of parallel profiles between CD (black) and SXD (red) for the case (a).} \label{fig_parallel_comp}
\end{figure}

Parallel profiles for the case (a) are shown in Fig. \ref{fig_parallel_comp}. 
The modification of the divertor leads to a transition of the SOL from the sheath-limited regime with flat $n_{\rm e}$ and $T_{\rm e}$ into the high-recycling regime. The temperatures in the divertor drop as the result of stronger collisional cooling and flux expansion giving rise to steeper $\nabla_{\parallel} T$. This is accompanied by an increase in the density (constant total pressure along the field line) and stronger ionization source in front of the divertor plate.
For the same pedestal density and input power into the SOL, the upstream $n_{\rm e}$ and $T_{\rm i}$ are unchanged, while the upstream $T_{\rm e}$ drops by several \% as $T_{\rm e}$ tends to flatten out more than  $T_{\rm i}$ due to larger parallel conductivity (subject of heat flux limiters). In the cases (b,c,d), all the upstream parameters $n_{\rm e}^{\rm u}$, $T_{\rm e}^{\rm u}$ and $T_{\rm i}^{\rm u}$ are comparable, see Tabs. \ref{table1} and \ref{table1_sxd}, and the reduction of the temperatures comes essentially from the divertor.

\subsection{Neutral species}
\label{sec_neut_sxd}

\begin{figure}[!h]
\scalebox{0.35}{\includegraphics[clip]{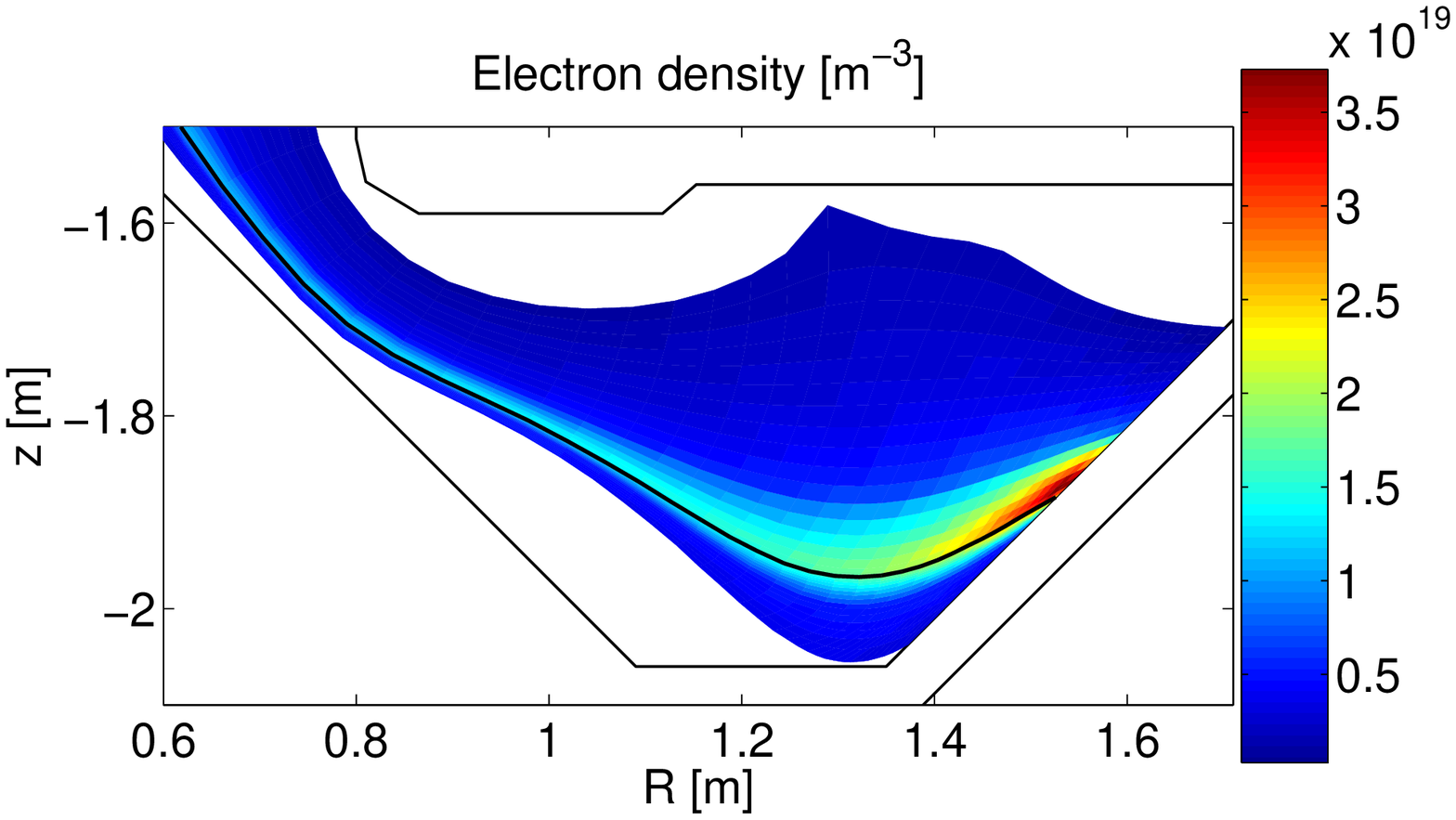}} 
\scalebox{0.35}{\includegraphics[clip]{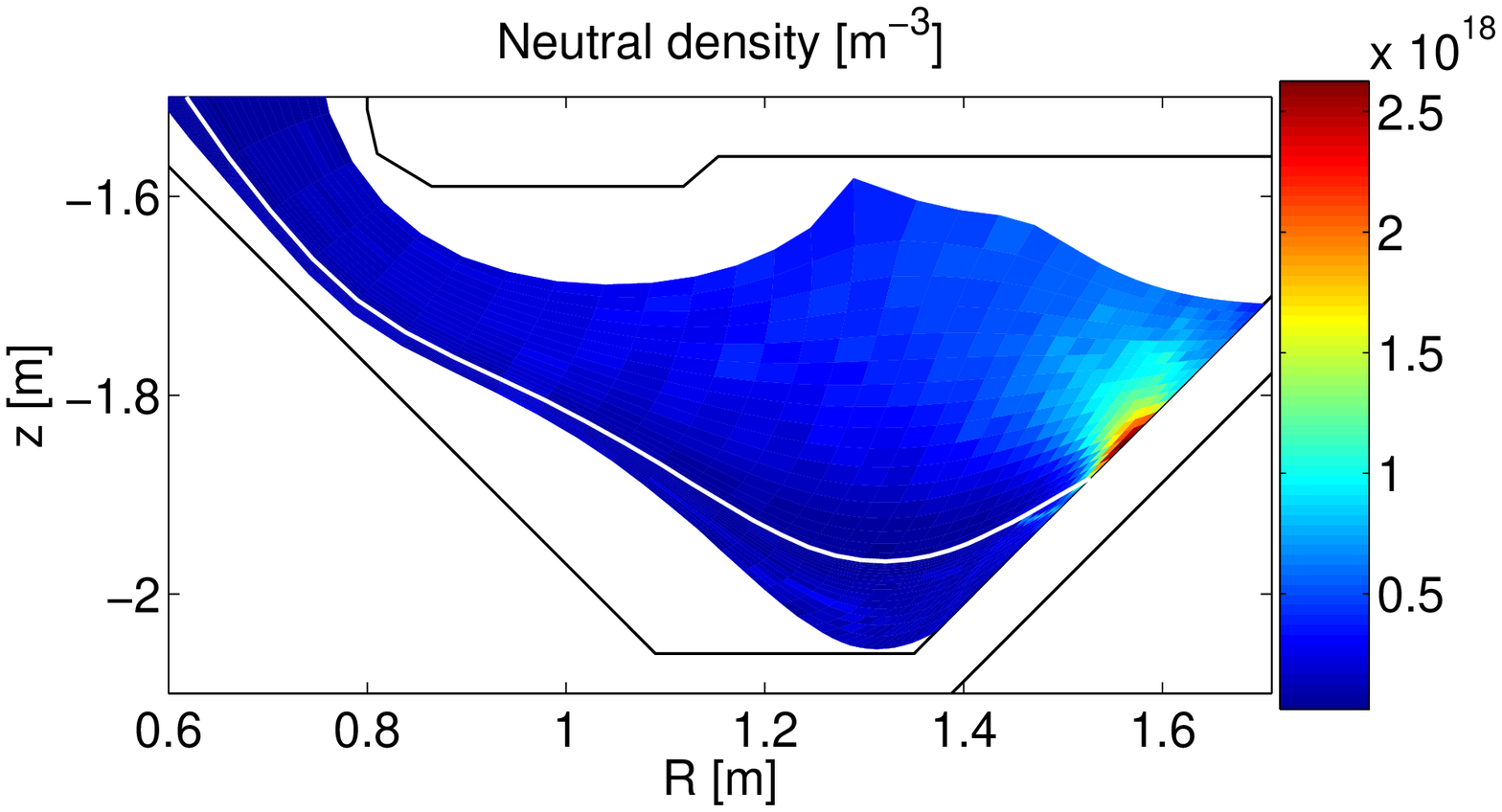}} \\
\scalebox{0.35}{\includegraphics[clip]{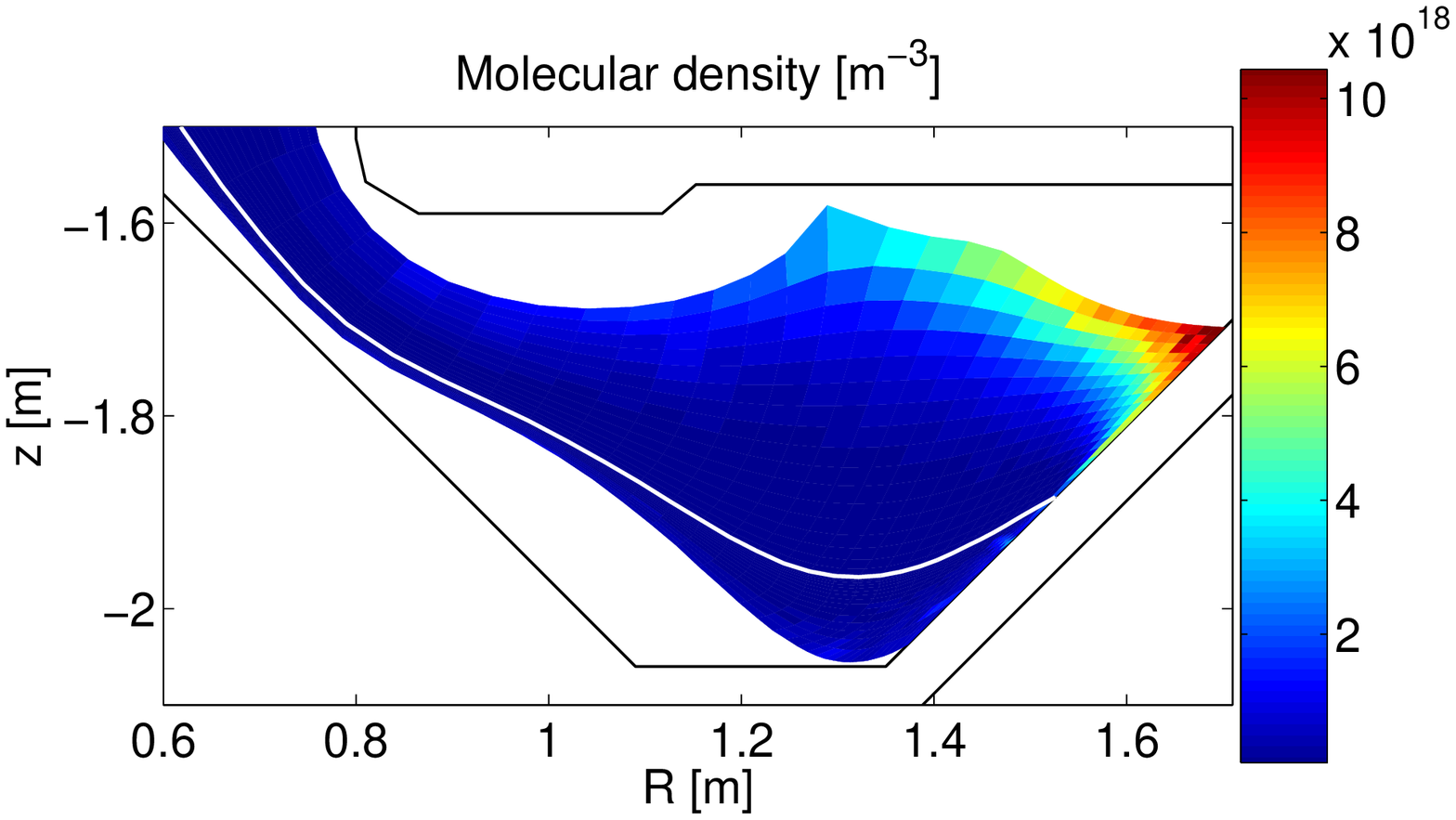}} 
\caption{The SXD configuration -- distribution of electrons, ${\rm D}_0$ and ${\rm D}_2$ in the divertor leg in the case (a). Compare with Fig. \ref{fig_distribution}.} \label{fig_distribution_sxd}
\end{figure}

\begin{figure}[!h]
\centerline{\scalebox{0.7}{\includegraphics[clip]{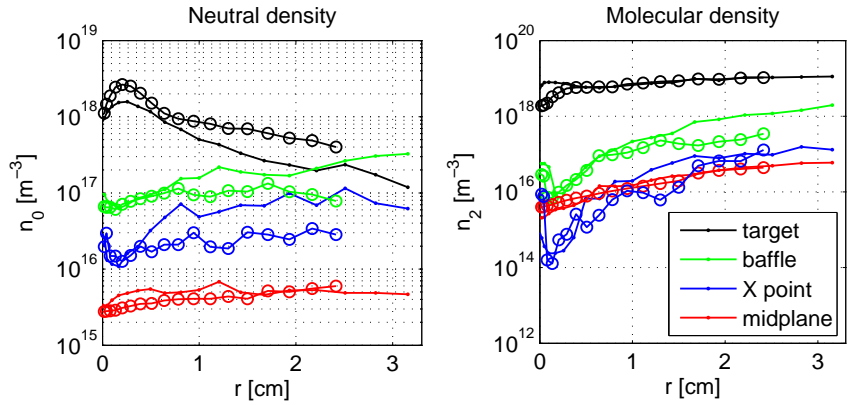}}} 
\caption{Radial profiles (mapped to the midplane) of the ${\rm D}_0$ and ${\rm D}_2$ density for the case (a) in CD (points) and SXD (open circles) at four poloidal locations in the outer SOL -- at the target (black), at the contact point of the plasma grid with the baffle (green), at the X-point (blue), at the midplane (red).} \label{fig_neutrals_sxd}
\end{figure}

Due to larger plasma densities in the divertor in SXD, more neutrals is recycled at the target and larger neutral densities are found in SXD (Fig. \ref{fig_distribution_sxd}) in comparison to CD (Fig. \ref{fig_distribution}). 
Neutrals species stay better separated from the main plasma compared to Fig. \ref{fig_distribution} where ${\rm D_0}$ and ${\rm D_2}$ particles reach further towards the X-point. 
Radial profiles of the ${\rm D_0}$ and ${\rm D_2}$ densities are shown in Fig. \ref{fig_neutrals_sxd} 
where the CD (points) and SXD (circles) H-mode cases are compared. The ${\rm D_0}$ densities are increased at the target compared to CD, while they are reduced at the location of the baffle nose, the X-point and at the midplane. The molecular densities in both configurations are comparable inside the simulation grid and they are larger at the midplane than at the X-point indicating that the molecules in the upstream SOL originate mainly from the main chamber (see also Tab. \ref{table_atomfluxes} shown below).

\begin{figure}[!h]
\centerline{\scalebox{0.5}{\includegraphics[clip]{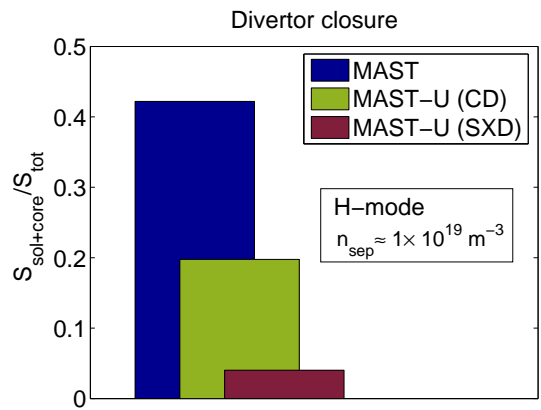}}} 
\caption{Divertor closure for the case (a) measured as the ionization source outside the divertor with respect to the total ionization source for three different divertor configurations -- MAST (figure in Tab. \ref{table_atomfluxes}), MAST-U with CD, MAST-U with SXD (Fig. \ref{fig_grid_sxd}).} \label{fig_closure}
\end{figure}

An improved divertor closure in SXD is demonstrated in Fig. \ref{fig_closure}, where the divertor closure is measured as the ionization source outside the divertor with respect to the total ionization source. The comparison is shown for H-mode plasma of $n_{\rm sep}\approx 1\times 10^{19}$ m$^{-3}$, the case (a), in three divertor configurations -- MAST, MAST-U with CD, MAST-U with SXD, but the result is similar for the cases (b,c,d). An improvement between MAST and MAST-U (a factor of 2) is caused by the baffle reducing the neutral and molecular fluxes from the divertor chamber to the main chamber. An additional improvement (a factor of 5) between CD and SXD is caused by larger collisionality in SXD (shorter mean free path and larger distance between the neutral source and the X-point) resulting in a reduced leakage of neutrals from the divertor to the upstream SOL.
\begin{table}[!h]
\begin{minipage}[c]{0.75\linewidth}
\begin{tabular}{l|rrr|rrr}
 & \multicolumn{3}{c}{atoms} & \multicolumn{3}{c}{molecules}  \\
 & MAST & CD & SXD & MAST & CD & SXD  \\
\hline
target & -2.77 & -1.58 & -10.34 & -2.11 & -3.35 & -6.06 \\
X-point & -0.65 & -0.90 & -0.41 & -0.22 & -0.12 & -0.01 \\
PFR & -0.24 & -0.42 & 0.05 & 0.29 & 0.33 & 0.41 \\
div. chamber & 4.37 & 9.67 & 16.54 & -0.47 & -4.34 & -7.52 \\
core & -3.62 & -1.27 & -0.85 & -0.74 & -0.11 & -0.06 \\
main chamber & 2.23 & 1.14 & 0.35 & -4.47 & -0.87 & -0.63 \\
\hline
\end{tabular}
\caption{Fluxes of ${\rm D_0}$ and ${\rm D_2}$ in [$10^{21}$ s$^{-1}$] for three divertor configurations - MAST (before upgrade, see the geometry on the right), CD and SXD (MAST-U). The fluxes are integrated across specific surfaces defined in the plot on the right, poloidal fluxes (target, X-point) are calculated in the direction of the $x$ coordinate and radial fluxes (core, PFR, main and divertor chambers) in the direction of the $y$ coordinate.}  
\label{table_atomfluxes}
\end{minipage}
\hspace{0cm}
\begin{minipage}[c]{0.20\linewidth}
\vspace{0cm}
\scalebox{0.53}{\includegraphics[clip]{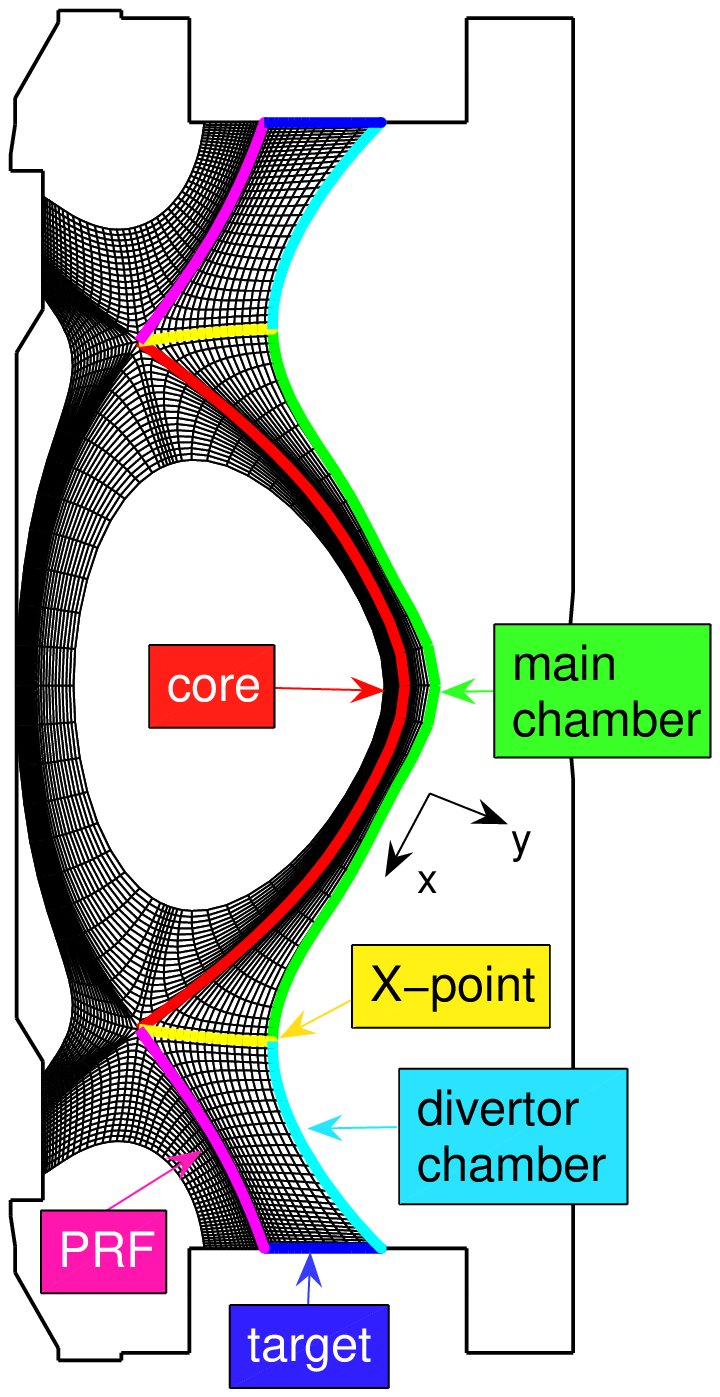}}
\end{minipage}
\end{table}

In order to confirm that the ionization source in the upstream SOL originates from neutrals penetrating from the divertor, Tab. \ref{table_atomfluxes} shows fluxes of ${\rm D_0}$ and ${\rm D_2}$ in the simulation across the outer lower divertor and the upstream SOL. 
In spite of large ${\rm D_0}$ flux from the target in SXD (one order of magnitude larger than in CD or MAST), the flux from the divertor to the upstream SOL across the X-point in SXD is by one order of magnitude smaller than in CD. Similarly, the molecular flux to the upstream SOL is reduced.

The molecular flux from the divertor to the upstream SOL is negligible in all configurations compared to the flux from the main chamber that is particularly large in MAST as there is no baffle, i.e. the divertor and main chambers are connected.

The atomic flux from the divertor region to the upstream SOL contributes largely to the flux to the core, as it is of the same magnitude. The flux to the core is clearly largest for MAST, both in terms of neutrals and molecules, and reduced in SXD (by a factor of 4 for ${\rm D}_{0}$ and a factor of 12 for ${\rm D}_{2}$ with respect to MAST). In MAST, the large atomic flux to the core results from the large escape of molecules flux from the divertor to the main chamber (no baffle) followed by dissociation in the upstream SOL. 

Note that neutral-neutral collisions are not included in the EIRENE version used here. Nevertheless, we do not expect the neutral-neutral collisions to play a role in the studied MAST cases where the neutral densities at the target do not reach $\sim 10^{20}$ m$^{-3}$ (Fig. \ref{fig_neutrals_sxd}).

\subsection{Power losses and radiation}
\label{sec_rad_sxd}

In the Super-X configuration, the plasma volume in the divertor is significantly increased. It is not obvious that this 
also implies larger radiation volume and that longer $L_{\parallel}$ actually enables more power to be removed. On one hand, arguments for larger power losses in SXD are the better divertor closure resulting in higher neutral pressures and reduced temperatures at which radiation is effective. On the other hand, steeper parallel temperature gradients associated with longer $L_{\parallel}$ could shorten the region that radiates, as neutral/impurity radiation takes place only in a certain temperature band. 

\begin{figure}[!h]
\centerline{\scalebox{0.4}{\includegraphics[clip]{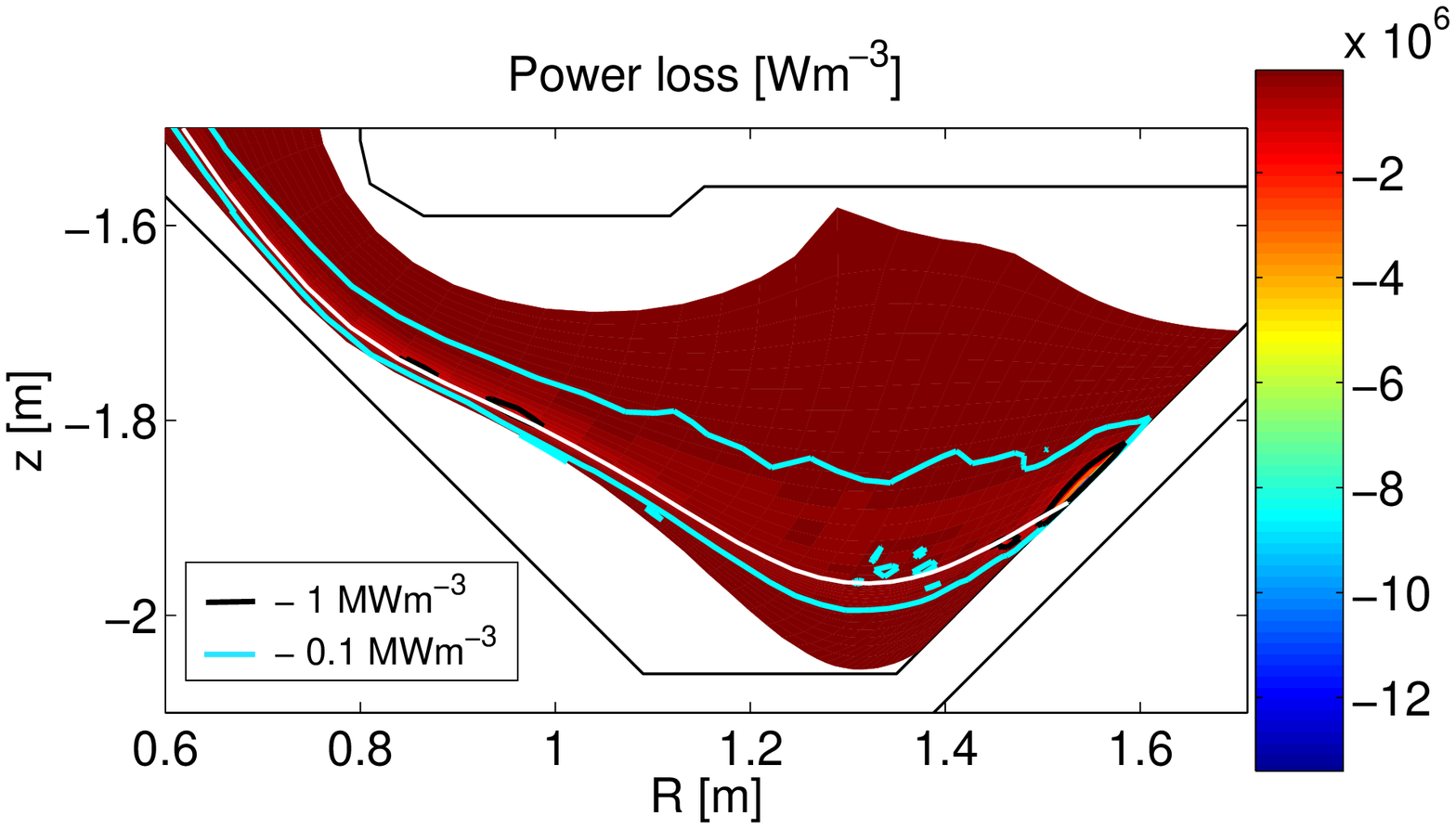}}
\scalebox{0.4}{\includegraphics[clip]{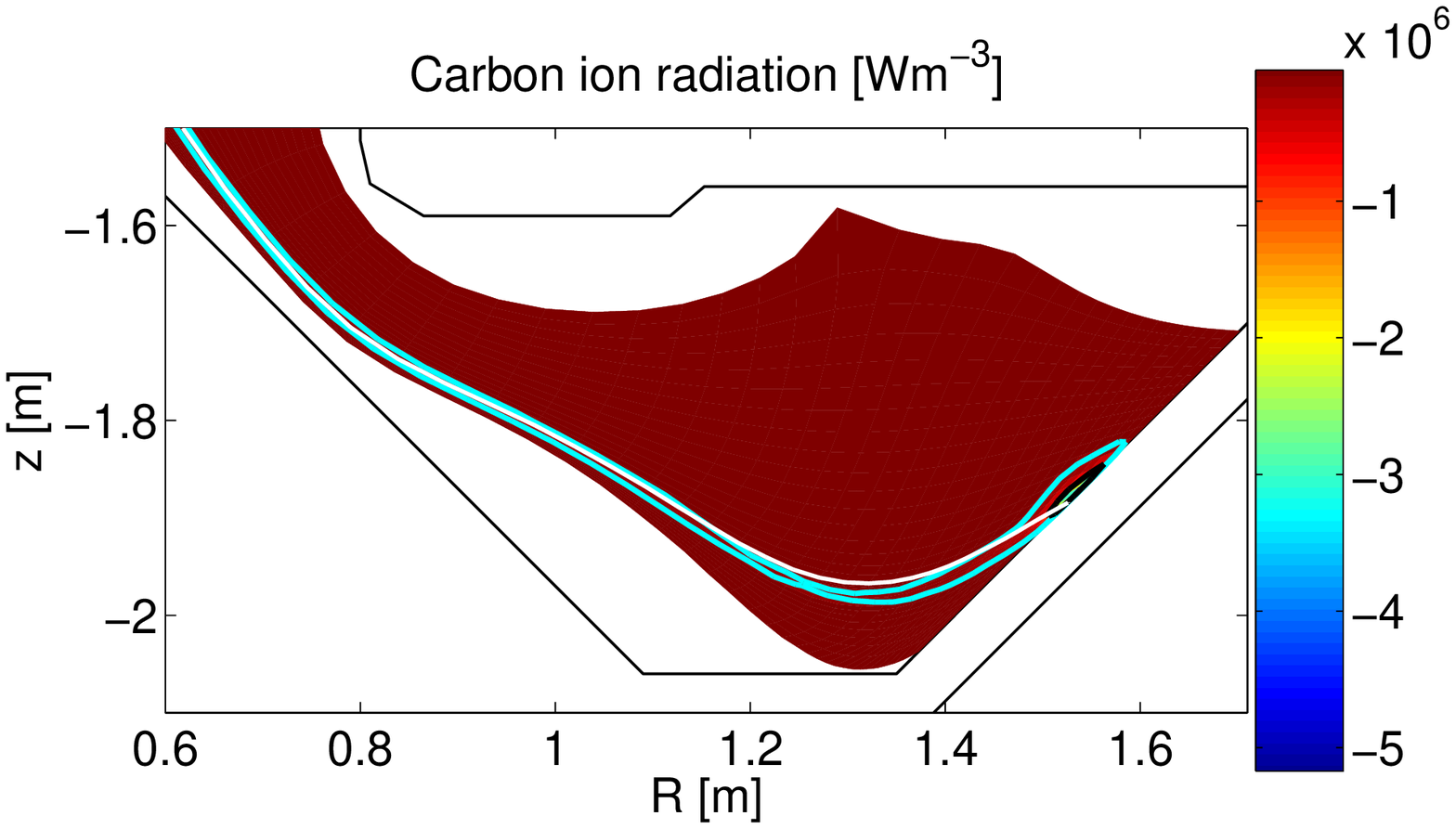}}
} 
\caption{The SXD configuration -- (left) distribution of total power losses for the case (a). The contour lines show regions in the divertor where the power loss exceeds a certain limit. (right) Distribution of carbon line radiation. Compare with Fig. \ref{fig_volumetric}.} 
\label{fig_volumetric_sxd}
\end{figure}

In order to compare SXD and CD in MAST-U, regions where the power loss exceeds a certain threshold are defined and shown in Figs. \ref{fig_volumetric} and \ref{fig_volumetric_sxd} by contour lines. The black contour line (the power loss larger than 1 MWm$^{-3}$) shows that the most intensive loss is located very close to the target plate in a region with smaller temperatures and larger neutral densities. This region does not broaden in volume in SXD, however, it is small enough to neglect substantial fraction of the power loss. The blue contour line in Fig. \ref{fig_volumetric_sxd} on the left (the power loss larger than 0.1 MWm$^{-3}$) specifies a region where the power loss reaches 83\% of the power loss in the divertor in SXD. The power loss rate is a magnitude smaller than in the black region, but it spreads over much larger surface, and extends all the way from the target towards the X-point. In Fig. \ref{fig_volumetric_sxd} right, only carbon ion line radiation is shown. The comparison with CD in Fig. \ref{fig_volumetric} for the case (a) demonstrates that both the total power loss and carbon radiation regions stretch with $L_{\parallel}$ and the trend is similar for the cases (b,c,d). 

\begin{figure}[!h]
\centerline{\scalebox{0.5}{\includegraphics[clip]{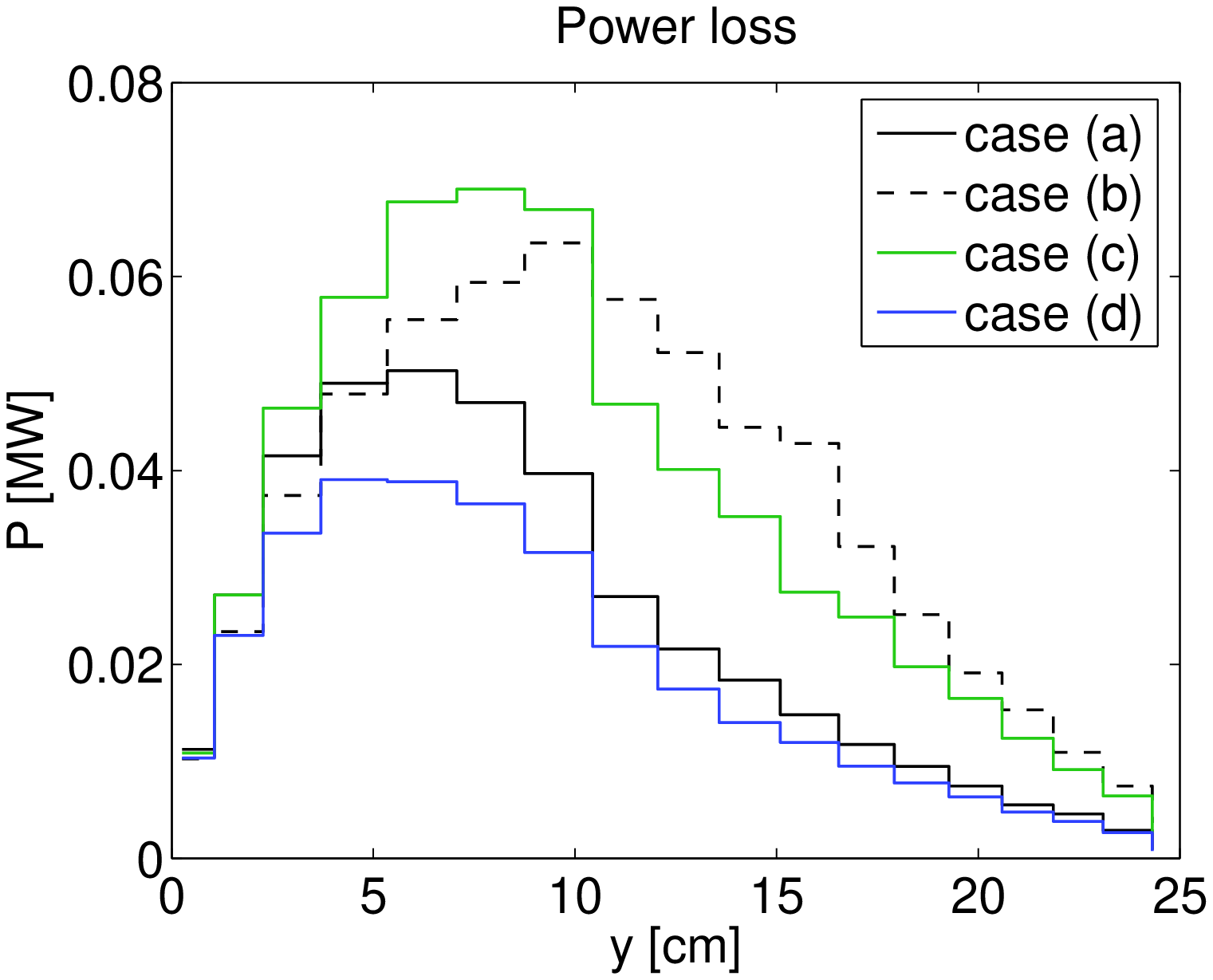}}
\scalebox{0.5}{\includegraphics[clip]{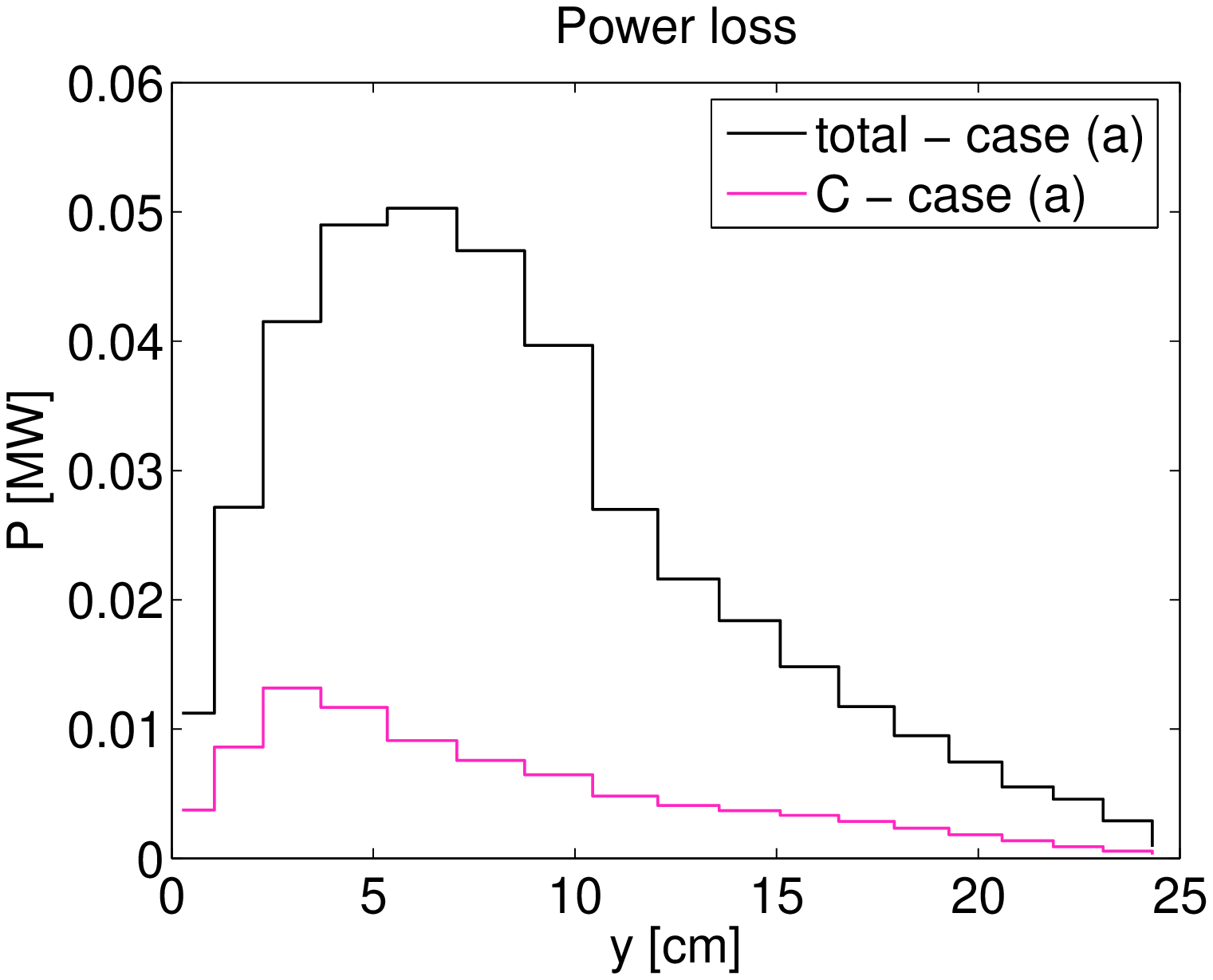}}
} 
\caption{The SXD configuration -- (left) volumetric power losses in the outer SOL calculated for each flux tube individually. All four cases are shown -- the H-mode case (a) (solid), the L-mode case (b) (dashed), the higher density case (c) (green) and the lower power case (d) (blue). (right) The total power loss and the loss due to carbon radiation are shown for the case (a). Compare with Fig. \ref{fig_volumetric2}.} \label{fig_volumetric_sxd2}
\end{figure}

In Fig. \ref{fig_volumetric_sxd2} left, the radial distribution of the power losses is shown for all studied cases. In Fig. \ref{fig_volumetric_sxd2} right, carbon radiation is separated for the case (a), again indicating that deuterium dominates the power removal. As in CD, increased density or wider SOL support radiation losses through increased collisionality. In both cases (b) and (c), the power loss is larger both in terms of the amplitude and the radial extend. In the low power case (d), the radiation pattern is similar to the one found in the case (a), however more power with respect to $P_{\rm inp}$ is removed. 

\begin{table}[!h]
\begin{center}
\begin{tabular}{|l|rrrr|rrrr|}
\hline
 & \multicolumn{4}{|c|}{total} & \multicolumn{4}{|c|}{carbon} \\
     & (a) & (b) & (c) & (d) & (a) & (b) & (c) & (d) \\
\hline
$P_{\rm vol}$ in the core & 4.7\% & 3.7\% & 1.8\% & 4.3\% & 0.8\% & 1.3\% & 1.0\% & 1.4\%\\
$P_{\rm vol}$ in the SOL & 28.9\% & 43.4\% & 41.1\% & 45.7\% & 8.1\% & 10.9\% & 9.6\% & 11.5\%\\
$P_{\rm vol}$ in the outer divertors & 25.5\% & 38.6\% & 37.5\% & 40.9\% & 6.1\% & 8.3\% & 7.3\% & 8.5\%\\
$P_{\rm vol}$ in the blue region & 21.2\% & 36.3\% & 34.3\% & 32.9\% & 3.3\% & 3.2\% & 3.8\% & 3.8\%\\
$S$ of the blue region [m$^2$] & 0.158 & 0.275 & 0.230 & 0.137 & 0.026 & 0.037 & 0.030 & 0.026\\
$V$ of the blue region [m$^3$] & 1.170 & 2.026 & 1.649 & 1.015 & 0.186 & 0.260 & 0.217 & 0.189\\
\hline
\end{tabular}
\caption{The SXD configuration -- the first four rows show the total volumetric power loss $P_{\rm vol}$ in terms of $P_{\rm inp}$ and the power loss due to carbon radiation separately in four SOLPS simulations (a), (b), (c) and (d) in different regions: (i) in the core (the closed field line part of the grid), (ii) in the SOL (the whole grid outside the separatrix), (iii) in the outer divertors (the region displayed in Fig. \ref{fig_volumetric}), (iv) in the blue region (defined in Fig. \ref{fig_volumetric} by the contour line). The bottom rows show the surface $S$ and the volume $V$ of this region. The total power loss is calculated from EIRENE as the total energy loss caused to the plasma due to all plasma-neutral interactions including neutral and impurity radiation, charge exchange, ionization. Compare with Tab. \ref{table_volumetric}.}
\label{table_volumetric_sxd}
\end{center}
\end{table}

This is summarized in Tab. \ref{table_volumetric_sxd} which also proves enhanced volumetric power losses in SXD in all cases compared to CD, as well as an expansion of the power loss area. The power loss caused by carbon ion radiation is separated and shows that the carbon radiation zone also expands in volume with increased $L_{\parallel}$ in SXD. In addition, the power radiated by carbon in the core is reduced in SXD with respect to CD (up to a factor of 0.6) at the same time as the power radiated in the SOL is increased (up to a factor of 2.3). Better impurity screening is achieved in SXD also in terms of carbon particle densities. Finally, the effect of carbon concentration has been tested by increasing the chemical sputtering yield from 1\% to 3\%. Stronger sputtering leads to $15-28$\% carbon radiation in the SOL, $2-3$ times more than the lower sputtering yield. This also means that the ability of SXD to radiate more than CD is slightly larger for higher sputtering yield (see also section \ref{sec_gb_sxd}).

\subsection{Particle and energy fluxes}

\subsubsection{Target fluxes}
\label{sec_sxd_tf}

The energy flux at the target in SXD can be reduced with respect to CD by the effects of geometry (magnetic flux expansion), by volumetric power losses along the field line and by a broadening of the target wetted area due to radial transport. The magnetic flux expansion will be discussed in section \ref{sec_fx_sxd} and the effect of radial transport in section \ref{sec_lpar_effect}.

\begin{figure}[!h]
\centerline{\scalebox{0.7}{\includegraphics[clip]{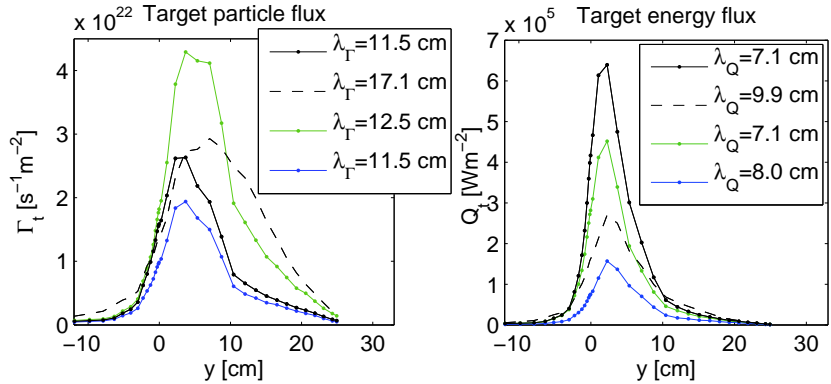}}
}
\caption{The SXD configuration -- particle (deuterium ions) and power (deuterium ions plus electrons) loads at the outer target as functions of the target coordinate $y$ for the case (a) (solid), case (b) (dashed), case (c) (green) and case (d) (blue). Compare with Fig. \ref{fig_fluxes}.} \label{fig_fluxes_sxd}
\end{figure}

Particle and energy fluxes deposited at the target in SXD are shown in Fig. \ref{fig_fluxes_sxd}. In the L-mode case (b), the target wetted area broadens and the peak $Q_{\rm t}$ drops with respect to the case (a). In the higher density case (c), $Q_{\rm t}$ drops at the same $P_{\rm inp}$ and $\lambda_Q$, as there is more power removed along the field line (this was not the case in CD in Fig. \ref{fig_fluxes} where radiation had smaller effect). In the case (d) where $P_{\rm inp}$ is a factor of 2 smaller, $Q_{\rm t}$ drops by a factor of 4 (more than in CD, again due to stronger power loss).

\begin{figure}[!h]
\scalebox{0.7}{\includegraphics[clip]{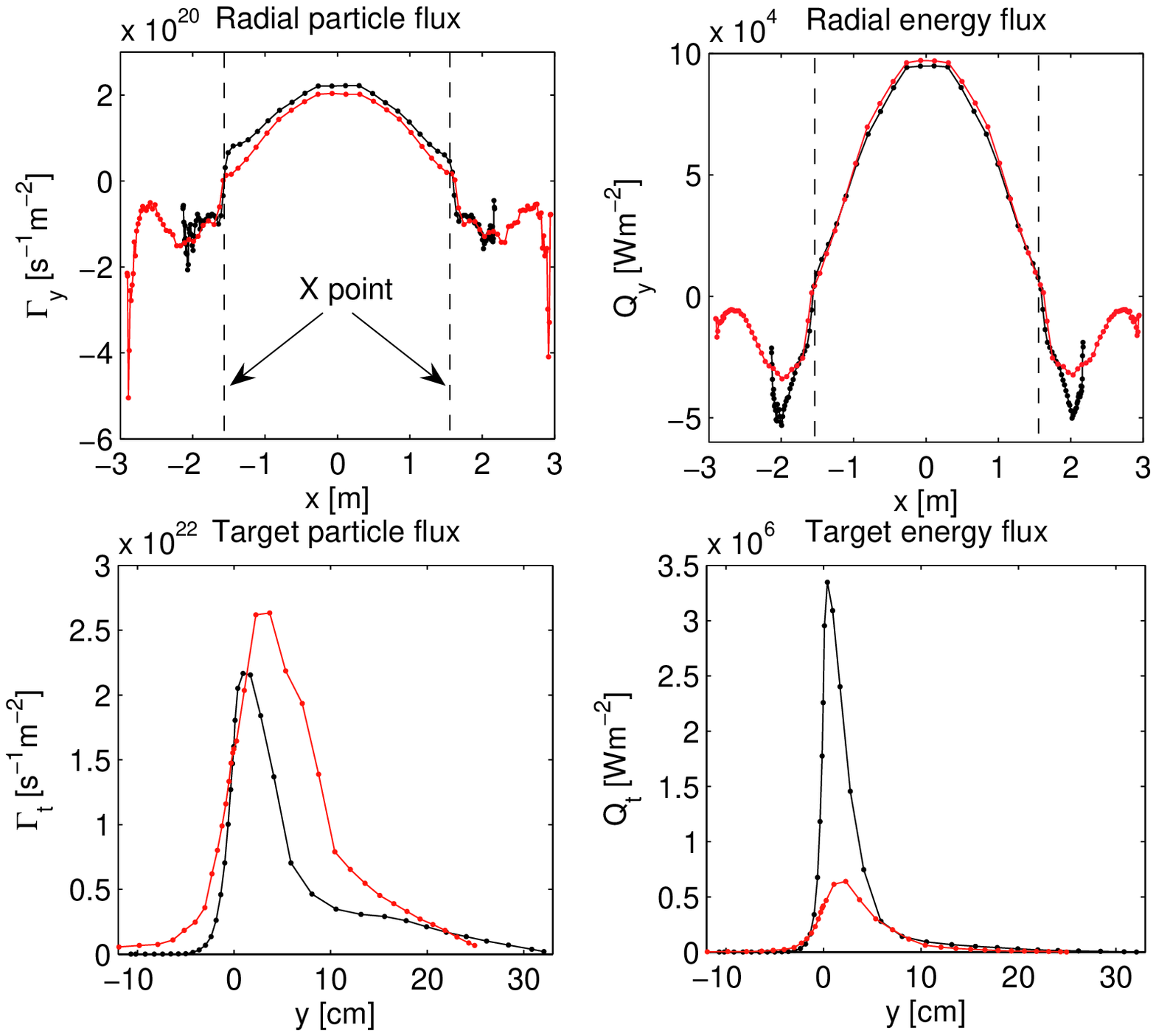}} \\
\scalebox{0.7}{\includegraphics[clip]{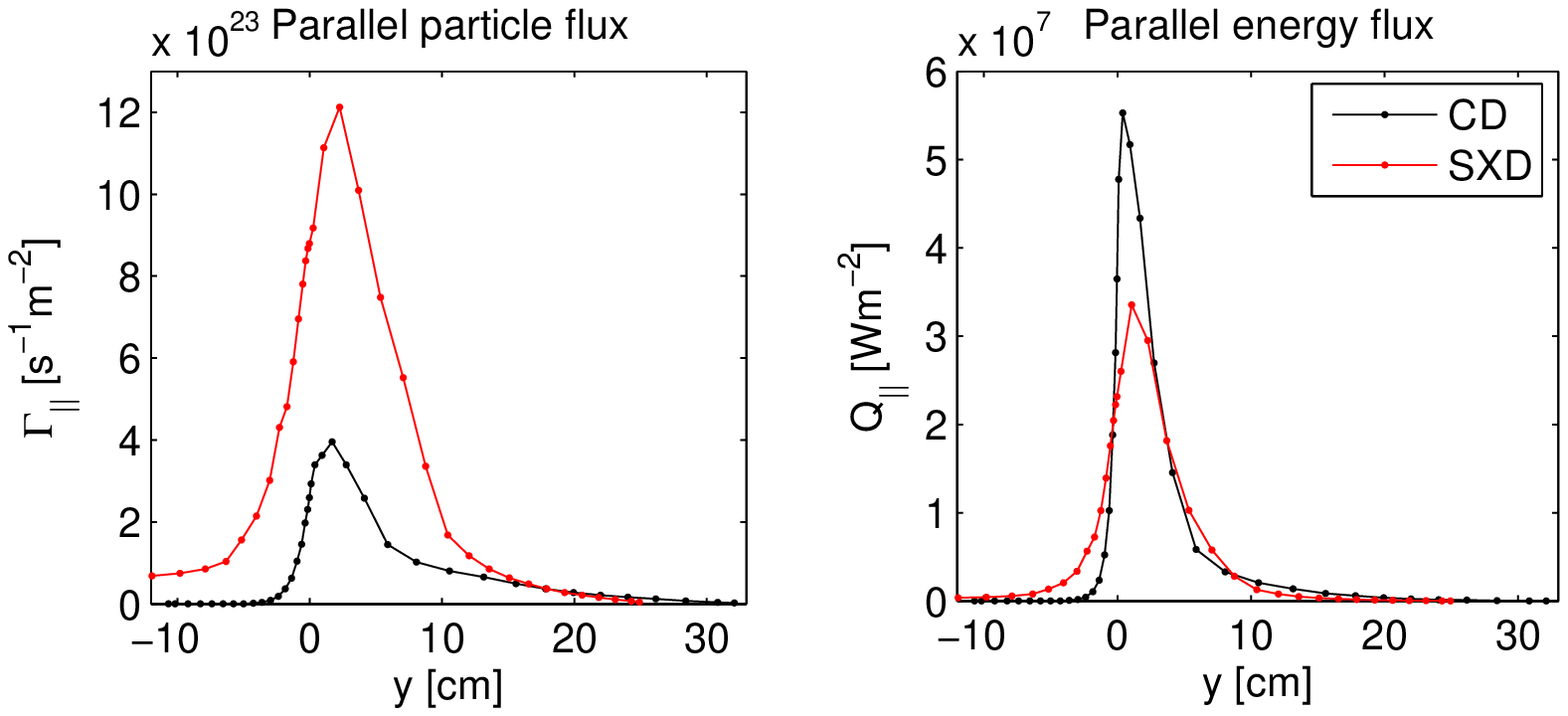}}
\caption{Direct comparison for CD and SXD in the case (a) -- (top) particle and energy fluxes crossing the separatrix at the outer side as functions of the poloidal coordinate $x$. (middle) Particle and power loads at the outer target as functions of the target coordinate $y$. (bottom) Parallel particle and energy fluxes at the outer target. The parallel fluxes are defined as: (i) the deuterium ion flux $\Gamma_{\parallel}=n_{\rm i}u_{\parallel}$, (ii) the electron energy flux $Q_{\rm \parallel,e}=5/2 n_{\rm e} u_{\parallel} k T_{\rm e}-\kappa_{\rm e}\partial (kT_{\rm e})/\partial s_{\parallel}$, (iii) the deuterium ion energy flux $Q_{\rm \parallel,i}=5/2 n_{\rm i} u_{\parallel} k T_{\rm i}+1/2m_{\rm i}n_{\rm i}u_{\parallel}^3-\kappa_{\rm i}\partial (kT_{\rm i})/\partial s_{\parallel}$.} \label{fig_fluxes_comp}
\end{figure}

By changing the divertor geometry from CD to SXD, the peak $\Gamma_{\rm t}$ does not change (larger flux expansion combined with stronger ionization source in the divertor), while the peak $Q_{\rm t}$ drops sharply (larger flux expansion  combined with larger volumetric power loss). This is demonstrated in Fig. \ref{fig_fluxes_comp} where a direct comparison between CD and SXD is presented for the case (a).

The energy flux width in the case (a) is $\lambda_Q^{\rm t}\approx 7.1$ cm which corresponds to $\lambda_Q^{\rm u}\approx 0.3$ cm (mapped to the midplane), and this is comparable to CD in section \ref{sec_cd_tf}. The particle flux width $\lambda_{\Gamma}^{\rm t}\approx 11.5$ cm mapped to the midplane as $\lambda_{\Gamma}^{\rm u}\approx 0.5$ cm is also similar to the value found in CD. For the cases (b,c,d), mapped to the midplane profiles have also the same widths, therefore the broadening of the energy flux at the target in SXD with respect to CD is solely caused by the magnetic flux expansion (section \ref{sec_fx_sxd}), while the radial diffusion has no effect (section \ref{sec_lpar_effect}) in the simulations presented here. This effect will be investigated further in the near future for broader range of input parameters. 

Fig. \ref{fig_fluxes_comp} at the top displays poloidal profiles of the particle and energy fluxes crossing the separatrix at the outer side. The input fluxes to the SOL between the X-points are the same in CD and SXD by prescribing the same $n_{\rm core}$ and $P_{\rm inp}$. In addition, there is a non-negligible transport of the plasma from the main SOL below the X-point to the private flux region. The longer connection length in SXD allows more particles and energy to diffuse to the private flux region. The sharp negative peak at the target in SXD in the particle flux profile is due to larger recycling source and target density in SXD.

In the middle row, the target particle and power loads are compared and the bottom row shows the same comparison in terms of the parallel fluxes. For the same flux into the SOL (top row), a factor of 5.2 reduction of the power load $Q_{\rm t}$ is achieved at the target in SXD in the case (a), and the peak parallel energy flux $Q_{\parallel}$ is reduced by a factor of 1.6. If we want to reduce the target power load to a similar level in the standard divertor, we would have reduce the input power by this factor, unless we increase volumetric power losses in the SOL (e.g. by increasing the density to achieve more radiating regimes or detachment, or by radiating through injected impurities). The field line angle at the target $\vartheta$ ($Q_{\rm t}=Q_{\parallel}{\rm sin}\vartheta$) is $\vartheta\approx 3.5^{\rm  o}$ at the separatrix in CD and drops to the critical value $\vartheta\approx 1^{\rm o}$ in SXD.

In the higher collisionality cases (b,c,d) with larger $\chi_{\perp}$, higher density or lower $P_{\rm inp}$, the $Q_{\rm t}$ drop increases up to a factor of 9.1, see Tab. \ref{table_qt_sxd} row (i), due to increased power losses. The effect of SXD on the reduction of $Q_{\rm t}$ also increases with the sputtering, see Tab. \ref{table_qt_sxd} row (ii), as the power fraction radiated by carbon increases slightly more when the sputtering coefficient is increased in SXD than for the same increase in CD (Tabs. \ref{table_volumetric} and \ref{table_volumetric_sxd}). With stronger sputtering, carbon radiates more and the temperatures in the divertor are reduced. This also leads to increased importance of other loss processes as well (e.g. charge-exchange). In the cases (b) and (d) in row (ii) where the $Q_{\rm t}$ drop between CD and SXD is largest, the detachment threshold is reached in the SXD simulations.

\begin{table}[!h]
\begin{center}
\begin{tabular}{llrrrr}
    &  & (a) & (b) & (c) & (d) \\
\hline
\multirow{2}{*}{$Q_{\rm t}$ drop} & (i) & 5.2 & 7.5 & 7.2 & 9.1 \\
 & (ii) & 5.8 & 18.2 & 9.4 & 19.5 \\
\hline
\end{tabular}
\caption{Reduction of the peak target energy flux in SXD with respect to CD in the simulations defined in Tab. \ref{table_runs}. The first row (i) is for the reference set of simulations with the chemical sputtering yield of 1\%. The second row (ii) is for cases with increased sputtering yield 3\% to explore the effect of sputtering.}
\label{table_qt_sxd}
\end{center}
\end{table}

\subsubsection{Flux expansion}
\label{sec_fx_sxd}

One of the benefits of the SXD geometry is a large flux expansion in the divertor region resulting in a broadening of the energy deposition profile and a reduction of its peak value. 
For configurations in Fig. \ref{fig_grid_sxd}, the total flux expansion is responsible for approximately a factor of 4 reduction of the peak energy flux out of the total reduction shown in Tab. \ref{table_qt_sxd}.

The flux expansion in CD and SXD at each radial position is shown in Fig. \ref{fig_fx} for each expansion factor separately (see Appendix for a definition). In Tab. \ref{table_fx_sxd}, the effective flux expansion is shown as an average over 7 cm ($\approx \lambda_Q^{\rm t}$), or alternatively over 11.5 cm ($\approx \lambda_{\Gamma}^{\rm t}$). The values can be compared with Tab. \ref{table_fx} for CD. The total effective flux expansion is a factor of 4.27 larger in SXD, mainly due to the toroidal flux expansion that accounts for a factor of 1.92 (the target at larger radius in SXD), and the poloidal flux expansion which is a factor of 1.98 larger (reduced poloidal magnetic field in SXD, see Fig. \ref{fig_grid_sxd}). 

\begin{figure}[!h]
\centerline{\scalebox{0.5}{\includegraphics[clip]{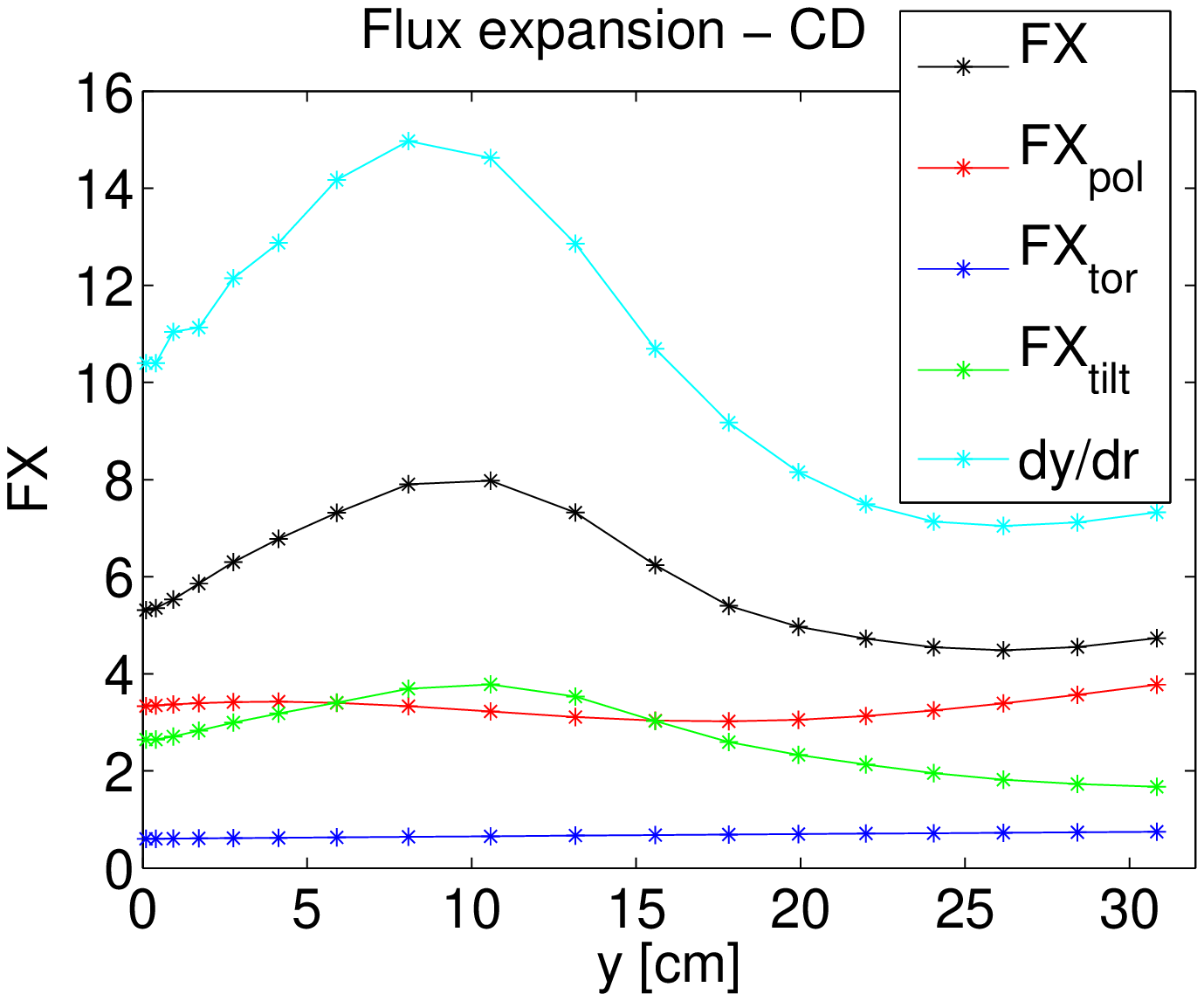}}
\scalebox{0.5}{\includegraphics[clip]{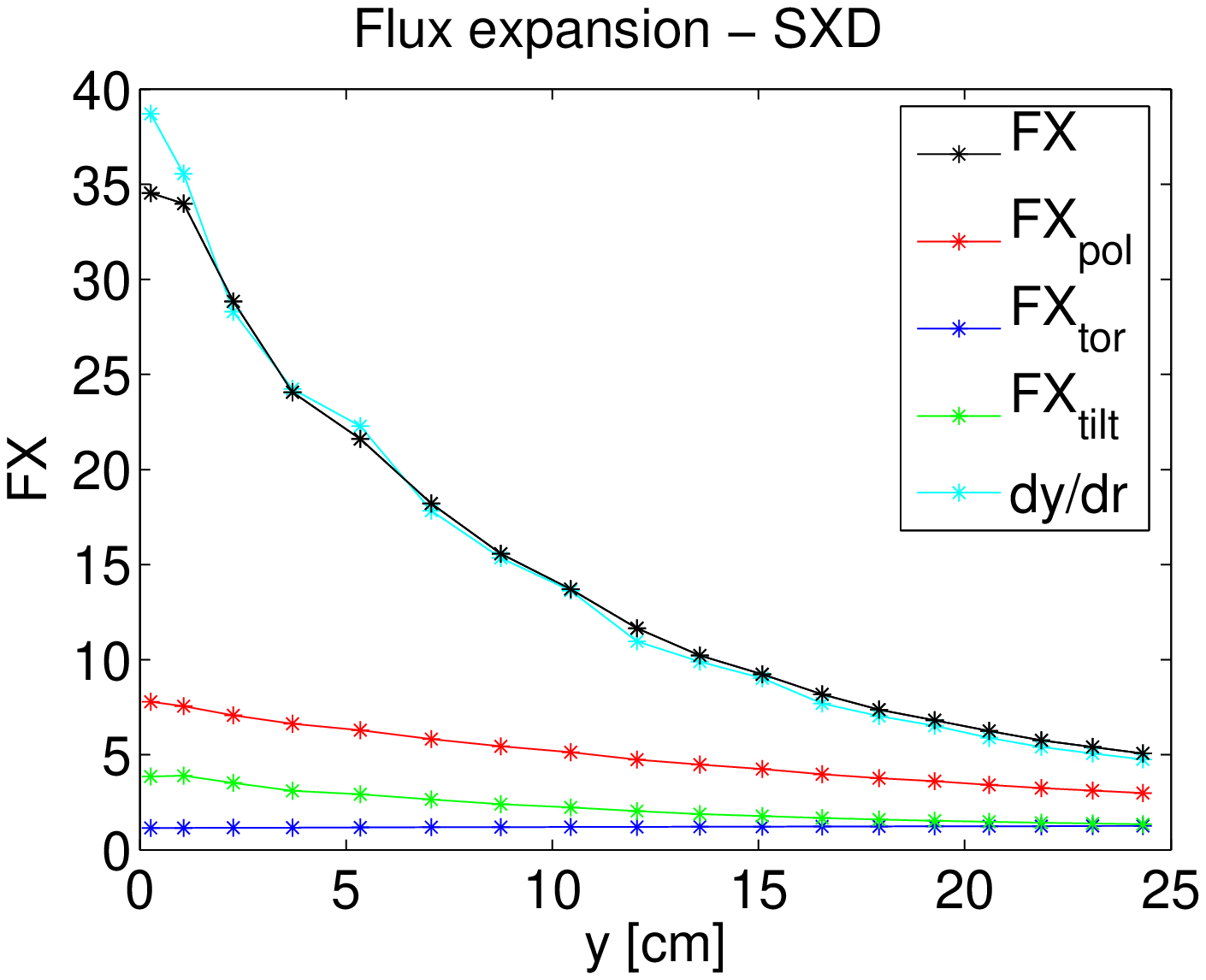}}
} 
\caption{Flux expansion factors in CD and SXD as functions of the target coordinate $y$. The total magnetic flux expansion $FX$ is separated into the poloidal flux expansion $FX_{\rm pol}$, the toroidal flux expansion $FX_{\rm tor}$ and the expansion due to the target tilting $FX_{\rm tilt}$. The expansion factor for mapping the target coordinate $y$ to the radial midplane coordinate $r$ is shown as ${\rm d}y/{\rm d}r$.} \label{fig_fx}
\end{figure}

\begin{table}[!h]
\begin{center}
\begin{tabular}{lrrrrrr}
 & $\langle FX\rangle$ & $\langle FX_{\rm tor}\rangle$ & $\langle FX_{\rm pol}\rangle$ & $\langle FX_{\rm tilt}\rangle$ & $\langle {\rm d}y/{\rm d}r\rangle$ \\
\hline
average over $\lambda_Q$ & 25.73 & 1.17 & 6.73 & 3.24 & 26.27 \\
average over $\lambda_{\Gamma}$ & 21.13 & 1.18 & 6.13 & 2.86 & 21.33 \\
\hline
\end{tabular}
\caption{
The SXD configuration -- flux expansion factors between the outer midplane and outer target. Compare with Tab. \ref{table_fx}. 
}
\label{table_fx_sxd}
\end{center}
\end{table}

The reduction of the target energy flux in the H-mode case (a) in Tab. \ref{table_qt_sxd} is not much larger than the reduction caused by the flux expansion, which can be interpreted in a way that it is mainly the magnetic topology that reduces $Q_{\rm t}$, while radiation has a smaller effect. In the cases (b,c,d) with larger collisionality, the volumetric power loss plays more important role with respect to the flux expansion (Tab. \ref{table_qt_sxd}), i.e. one benefits from SXD more in high density or detached regimes with stronger radiation (discussed in \cite{Eva3}).

Two comments should be made regarding the reduction of $Q_{\rm t}$ in SXD. In the case (a), the reduction of 5.2 seems to be a consequence of the flux expansion only (a factor of 4.27), in spite of larger radiated fraction in SXD (see Tabs. \ref{table_volumetric} and \ref{table_volumetric_sxd}). One of the reasons is a displacement between 
the energy flux and radiation profiles (Figs. \ref{fig_volumetric_sxd2} and \ref{fig_fluxes_sxd}).
Second, the power loss along the field line at the location of the peak $Q_{\rm t}$ is masked by stronger parallel heat transport in SXD and consequent increase of $Q_{\parallel}$ in the near SOL driven by larger $\nabla_{\parallel} T$ in SXD. This has been analyzed in \cite{Eva2} and will be discussed also in the next section.

\subsubsection{The role of $L_{\parallel}$ in SOL broadening}
\label{sec_lpar_effect}

Another geometric feature of SXD is a long connection length in the divertor (Fig. \ref{fig_grid_sxd}). It is expected that longer $L_{\parallel}$ leads to a SOL broadening in the divertor as a consequence of the competition between parallel and radial transport (longer distance to the target gives more time to spread radially).
However in the simulation, the SOL width in terms of $\lambda_Q$ and $\lambda_{\Gamma}$ is the same in CD and SXD when mapped to the midplane (sections \ref{sec_cd_tf} and \ref{sec_sxd_tf}) and there is also no obvious impact of the longer divertor on radial locations of the peak target values of the density and temperature. 

Let us therefore estimate the way $L_{\parallel}$ affects parallel losses. 
For particles, the parallel loss time can be defined as 
\begin{equation}
\tau_{\parallel}^n\equiv \frac{n}{\nabla_{\parallel}(nu_{\parallel})} \label{eq_loss1}
\end{equation}
using the continuity equation and assuming steady state. Eq. (\ref{eq_loss1}) results in
$\tau_{\parallel}^n=0.68$ ms for CD and $\tau_{\parallel}^n=1.7$ ms for SXD (calculated from values at the outboard midplane separatrix). Alternatively, $\tau_{\parallel}$ can be written as a function of $L_{\parallel}$ as $\tau_{\parallel}^n \approx L_{\parallel}/c_{\rm s}$ if the approximation $\nabla_{\parallel}\approx 1/L_{\parallel}$ is used, leading to $\tau_{\parallel}^n\approx 0.16$ ms for CD and $\tau_{\parallel}^n \approx 0.32$ ms for SXD. In both cases, longer time is needed in SXD for particles to reach the target (approximately a factor of 2), if we assume the same radial speed. The last assumption is justified by the usage of the same $D_{\perp}$ and fairly identical radial density profiles (Fig. \ref{fig_radial_comp_outer}).

\begin{figure}[!h]
\centerline{\scalebox{0.8}{\includegraphics[clip]{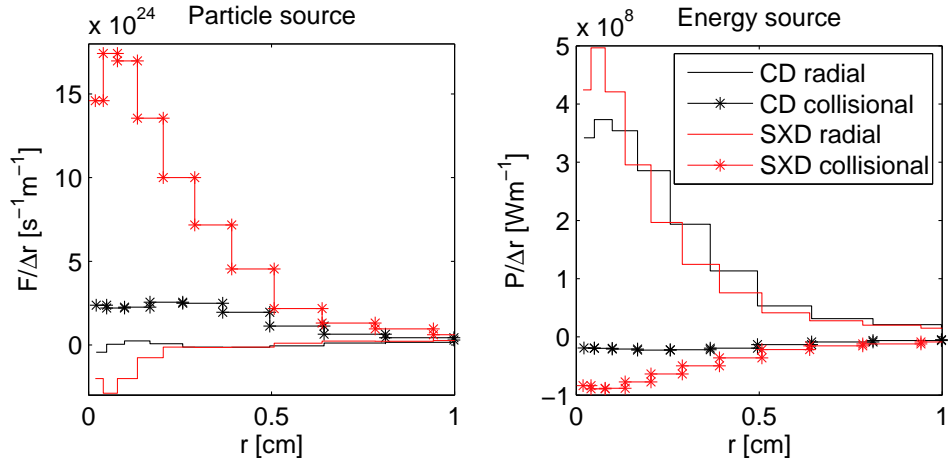}}} 
\caption{Magnitude of radial and collisional sources across the SOL. (left) Particle sources in each flux tube (displayed as the total integral source in the flux tube in [s$^{-1}$] divided by the flux tube width $\Delta r$) caused by radial transport and by recycling are shown for CD and SXD. For comparison, the profiles are mapped to the midplane coordinate $r$. (right) Energy sources separated as radial and collisional for each flux tube are shown. The gain of particles due to the radial transport is negligible in comparison to the ionization source in the divertor. In SXD, the radial source is negative in the near SOL, which means a particle sink from the flux tube to the private flux region).
The energy transport is dominated by the radial fluxes, however, there is a larger energy loss in SXD due to cooling in the divertor. 
} \label{fig_rad_coll}
\end{figure}

For particles, however, it is not the transport from the upstream SOL and core that dominates, but the ionization source in the vicinity of the target (see Fig. \ref{fig_rad_coll} left). 
The $\Gamma_{\rm t}$ profile is correlated with the ionization source and the distribution of this source is a function of the atomic and plasma density, but also the plasma temperature. In addition to plasma-neutral interactions in the divertor, one can notice a non-negligible transport from the divertor region into the private flux region across the separatrix (Figs. \ref{fig_rad_coll} and \ref{fig_fluxes_comp} top), which can also contribute to the shape of the deposition profile. 

For energy, the deposition profile is mainly determined by the power arriving from the upstream SOL and core (Fig. \ref{fig_rad_coll} right). In the simulation, the transport process along the field line is a combination of convection and conduction, although conduction exceeds convection by 1--2 orders of magnitude, and electron and ion conductive losses are of comparable importance. A simple and often used approximation of the parallel loss time for energy $\tau_{\parallel}^E \approx 3nL_{\parallel}^2/2\kappa_{\rm e}$ would result in $\tau_{\parallel}^E \approx 3.5$ $\mu$s for CD and $\tau_{\parallel}^E \approx 27.2$ $\mu$s for SXD and the longer $\tau_{\parallel}^E$ in SXD would indeed suggest a substantially wider SOL in terms of $\lambda_Q$. However, the SOLPS results show that unlike for the convection loss time, which can be approximated as $\tau_{\parallel}^n \propto \frac{L_{\parallel}}{c_{\rm s}}$, the simple estimate above for $\tau_{\parallel}^E$ does not hold, as it appears we can not approximate the gradient length scale by $L_{\parallel}$ for the temperature. More general form for $\tau_{\parallel}^E$ can be obtained using the energy equation and assuming steady state and conduction-dominated transport:
\begin{equation}
\tau_{\parallel}^E\equiv \frac{\frac{3}{2}n(T_{\rm e}+T_{\rm i})}{-\kappa_{\rm e}\nabla_{\parallel}^2 T_{\rm e}-\kappa_{\rm i}\nabla_{\parallel}^2 T_{\rm i}}. \label{tau_E}
\end{equation}
From Eq. (\ref{tau_E}), one obtains $\tau_{\parallel}^E \approx 3.8$ $\mu$s for CD and $\tau_{\parallel}^E \approx 2.7$ $\mu$s for SXD, i.e. comparable transport times in both geometries. In Eq. (\ref{tau_E}), the parallel gradient length for the temperature is shortened in SXD opposed to the simple approximation above where the gradient length grows with $L_{\parallel}$. 

To conclude this section, the absence of an additional power and particle flux broadening in SXD in these simulations is explained by processes of the recycling (for particles) and the cooling (for energy), both affecting the parallel transport and playing a role in shaping the $\Gamma_{\rm t}$ and $Q_{\rm t}$ profiles. For energy, cooler plasma in the divertor in SXD leads to steeper parallel temperature gradients which drive stronger parallel losses cancelling the effect of longer $L_{\parallel}$. It is a different result than expected and it is a consequence of different regimes in CD and SXD (SXD achieves cooling while CD does not). A detached regime where both CD and SXD divertor plasmas are cold should be investigated, as in such case the difference in the parallel gradient length between CD and SXD is minimized. It is expected that the SXD advantages in such case would be stronger.

\subsubsection{Global balance}
\label{sec_gb_sxd}

As for CD in Tab. \ref{table2}, global power and particle balance is calculated for the SXD simulations in Tab. \ref{table2_sxd}. Similar trends are found in SXD as in Tab. \ref{table2} -- the volumetric power losses increase with larger radial transport and larger densities reducing the total power received by the targets. The particle and energy fluxes deposited at the outer wall are negligible compared to the target fluxes, even though little more power escapes into the private flux region as the divertor leg gets longer. More importantly, compared with CD, the power deposited at the outer targets drops for the same power into the SOL by approximately a factor of 2 in the cases (b,c,d) and 1.5 for the case (a), and the particle load at the target is larger. In all cases, approximately twice as much power is removed from the plasma before reaching the target in SXD in comparison to CD, in these baseline simulations without any additional impurity species. 

The volumetric power losses are caused by a combination of processes including impurity radiation, neutral radiation and all plasma-neutral energy exchange processes. The effect of carbon impurity is investigated separately. The power fraction radiated by carbon in SXD increases from $9-13$\% for the sputtering yield of 1\% to $17-30$\% for the sputtering of 3\%. For the case with higher carbon concentration, the total power loss in SXD ($42-73$\% of $P_{\rm inp}$) is again approximately twice larger than in CD ($19-34$\% of $P_{\rm inp}$), but the power to the outer targets is reduced more between CD and SXD than for lower carbon concentration (from $P_{\rm t,out}\approx 56-72$\% in CD to $14-46$\% in SXD). Therefore the effect of SXD on the reduction of $P_{\rm t,out}$ increases with increased carbon content.  

\begin{table}[!h]
\begin{center}
\begin{tabular}{lcccc}
    & (a) & (b) & (c) & (d) \\
\hline
$P_{\rm sol,in}$, $P_{\rm sol,out}$ &  10.0\%, 85.0\% & 11.8\%, 83.9\% & 11.8\%, 85.6\% & 11.1\%, 84.5\% \\
$P_{\rm t,in}$, $P_{\rm t,out}$ &  10.9\%, 52.7\% & 11.8\%, 31.6\% & 11.8\%, 37.2\% & 10.9\%, 29.9\% \\
$P_{\rm wall}$, $P_{\rm pfr}$ & 1.5\%, 1.6\%  & 3.1\%, 1.8\% & 2.5\%, 1.7\% & 2.3\%, 2.0\% \\
$P_{\rm vol}$ &  35.6\% & 53.7\% & 48.9\% & 56.9\% \\
$P_{\rm vis}$ &  2.4\% & 2.0\% & 2.1\% & 2.1\% \\
\hline
$F_{\rm sol}$ & $3.7\times 10^{21}$ s$^{-1}$ & $7.8\times 10^{21}$ s$^{-1}$ & $7.2\times 10^{21}$ s$^{-1}$ & $3.9\times 10^{21}$ s$^{-1}$\\
$F_{\rm sol,out}$ & 81.1\% & 83.2\% & 82.5\% & 81.9\%  \\
$F_{\rm t,out}$ & $6.3\times 10^{22}$ s$^{-1}$ & $10.0\times 10^{22}$ s$^{-1}$ & $10.9\times 10^{22}$ s$^{-1}$ & $4.5\times 10^{22}$ s$^{-1}$  \\
$F_{\rm wall,out}$ & 1.4\% & 1.9\% & 1.5\% & 1.7\% \\
\hline
\end{tabular}
\caption{The SXD configuration -- power and particle balance. Compare with Tab. \ref{table2} where also all the quantities are defined.}
\label{table2_sxd}
\end{center}
\end{table}

\section{Conclusions}

The 2D SOL modelling code SOLPS has been used for a detailed comparison of example conventional and Super-X divertor plasmas in MAST. Although it is known that standard SOL transport codes encounter problems to reproduce high density experiments and have limitations in the physics they presently include, the codes are powerful for comparative studies where they can capture essential trends. The picture of the SOL transport is simplified in the sense that laminar transport is assumed on a constant-in-time magnetic equilibrium, radial transport is fixed and constrained typically as diffusive, kinetic and MHD effects are neglected, as is non-diffusive transport due to 3D turbulence (which would affect the value and nature of the SOL width). This approach limits the predictive capability of the codes (e.g. to determine the SOL width), but can be used to study parallel plasma transport from the core along the field lines to the divertor. The codes have special strength in coupling with Monte Carlo description of neutral transport which allows to capture essential atomic physics playing an important role in the divertor modelling. Neutral-neutral collisions are not taken into account, but we do not expect them to have a significant effect in the studied cases as the input power and the neutral pressure in the divertor are not high enough.

Two divertor geometries -- the conventional divertor (CD) and the Super-X divertor (SXD) -- are confronted in conditions of an inter-ELM H-mode MAST experiment with similar conditions in the core and upstream SOL. Four baseline simulations with different density, input power and radial transport are considered for each configuration, all with attached and non seeded plasmas and similar range of upstream parameters in CD and SXD (the temperatures of 62--114 eV in CD and 58--81 eV in SXD, the densities of 0.9--1.7 $\times 10^{19}$ m$^{-3}$). As simulated by SOLPS, the SXD configuration of MAST-U proves to significantly reduce the target energy fluxes even for attached hot plasma (a factor of 5.2--9.1 in our simulations compared to a factor of 4.3 caused by flux expansion). The target temperatures drop, allowing an earlier transition to the detachment (here by a factor of 3.2--7.1 from the sheath-limited SOL in CD with $44-108$ eV to the high-recycling SOL with $7-27$ eV in SXD). The effect of SXD on reducing the peak target energy flux is larger in regimes with higher density and wider SOL in which larger volumetric power loss is achieved. As continuation, a comparison at high densities and detached regimes is presented in \cite{Eva3}. Note that drift effects have not been tested in the frame of this paper, however, they have been studied for MAST separately in \cite{Vladimir1}. The key role is played by the poloidal $\rm{E \times B}$ drift, which in a connected double null configuration results in an asymmetry of the plasma
parameters in the top and bottom divertors. In \cite{Vladimir1}, the change in the electron temperature varies between $0–-50$\% at the outer targets and this can for example affect the divertor detachment.

The divertor closure with respect to neutrals is increased in MAST Upgrade thanks to the separation of the upper SOL and divertor regions by a baffle and in SXD with respect to CD thanks to the shorter ionization mean free path of the recycled neutrals and longer distance between the neutral source and the X-point. The ratio between the ionization source outside the divertor and the total ionization source is reduced by a factor of 10 compared to the standard MAST divertor. As the result, the flux of neutral species to the core is reduced, here by a factor of 4 for deuterium atoms and a factor of 12 for molecules. The divertor in SXD is also more closed for impurities, as the power radiated by carbon in SXD is increased in the SOL, while reduced in the core. 

In SXD with twice longer connection length, approximately twice more power is removed from the plasma in the SOL volume before reaching the target by radiation and collisional processes (mainly charge exchange). The region from which the power loss occurs expands with increased $L_{\parallel}$ in the direction of magnetic field lines. 
The effect of SXD on reducing the power deposited at the target and the peak energy flux at the target also seems to increase with increased carbon sputtering yield. 

For the same $D_{\perp}$ and $\chi_{\perp}$ in CD and SXD, the SOL width at the midplane is the same both in terms of $n$ and $T$, while $\lambda_Q$ and $\lambda_{\Gamma}$ broaden at the target in SXD thanks to the magnetic topology. Opposed to expectations, $\lambda_Q$ mapped to the midplane is comparable in CD and SXD and no obvious $\lambda_Q$ broadening is found with increased $L_{\parallel}$ as the result of radial diffusion. The reason is cooler plasma in the divertor in SXD in these cases and conduction dominated parallel heat transport, thanks to which a steeper $\nabla_{\parallel} T$ in SXD leads to stronger parallel losses cancelling the effect of the longer parallel distance. If CD had a similar target temperature, the SXD advantages would be expected to be stronger. From the simulation, it also appears that the parallel loss time for energy can not be approximated by the often used expression $\tau_{\parallel}^E \approx 3nL_{\parallel}^2/2\kappa_{\rm e}$ which misrepresents the dependence on $L_{\parallel}$.  

In summary, this study has shown in detail where the SXD and CD differ and thus can guide future research to optimise the benefits of long-leg divertor configurations. 

\section*{Acknowledgments}
This work was part-funded by the RCUK Energy Programme (grant number EP/I501045) and the European Communities under the contract of Association between EURATOM and CCFE. To obtain further information on the data and models underlying this paper please contact PublicationsManager@ccfe.ac.uk. The views and opinions expressed herein do not necessarily reflect those of the European Commission. The author acknowledges the support of the SOLPS team, especially D. Coster, X. Bonnin, V. Rozhansky, D. Reiter.

\clearpage
\section*{Appendix}

\subsection*{Flux expansion in conventional and Super-X divertors}
\label{sec_app1}

Flux expansion between the midplane and target, relating the target wetted area to the area in the upstream SOL, 
\begin{equation}
FX\equiv \frac{2\pi R^u}{2\pi R^{\rm t}} \frac{{\rm d}y}{{\rm d}r}=\frac{R^{\rm t}}{R^{\rm u}}\frac{B_x^{\rm u}B^{\rm t}}{B^{\rm u}B_x^{\rm t}}\frac{1}{{\rm sin}\beta} \label{eq_expansion}
\end{equation}
consists of the toroidal flux expansion $FX_{\rm tor}$ (related to the $R$ coordinate, i.e. the toroidal magnetic field $B_z\approx B$), the poloidal flux expansion $FX_{\rm pol}$ (related to the poloidal magnetic field $B_x$) and the expansion due to the target plate tilting $FX_{\rm tilt}$,
\begin{equation}
FX_{\rm tor}\equiv \frac{R^{\rm t}}{R^{\rm u}}, \quad FX_{\rm pol}\equiv \frac{B_x^{\rm u}B^{\rm t}}{B^{\rm u}B_x^{\rm t}}, \quad FX_{\rm tilt}\equiv \frac{1}{{\rm sin}\beta}.
\end{equation}
Here, $y$ is the target coordinate, $r$ is the radial coordinate at the midplane, $\beta$ is the angle between the field line and the target in the poloidal plane, $B_x/B$ is the pitch angle. In the poloidal plane, the flux tube width expansion between the upstream and target locations can be expressed in terms of the poloidal flux expansion
\begin{equation} 
\frac{{\rm d}y}{{\rm d}r}=\frac{B_x^{\rm u}B^{\rm t}}{B^{\rm u}B_x^{\rm t}}\frac{1}{{\rm sin}\beta}
\end{equation} 
using the condition of the constant poloidal magnetic flux on a flux tube $h B_x/B={\rm const}$ ($h$ is the flux tube width and we assume $B\approx B_z \propto 1/R$). 

The reduction of the energy flux between CD (index 1) and SXD (index 2) due to geometry effects can be estimated from a comparison of the target wetted areas. The parallel energy flux for the same input power changes due to the flux expansion as
\begin{equation}
\frac{Q_{\rm \parallel,1}}{Q_{\rm \parallel,2}}=\frac{A_{\rm \parallel,2}^{\rm t}}{A_{\rm \parallel,1}^{\rm t}}
\end{equation}
where $A_{\parallel}^{\rm t}=2\pi R^{\rm t}\lambda^{\rm t}(B_x/B)^{\rm t}$ is the wetted area at the target normal to the parallel direction. Using $\lambda^{\rm t}(B_x/B)^{\rm t}=\lambda^{\rm u}(B_x/B)^{\rm u}$, the area can be expressed in upstream quantities. 
If we assume the same conditions in the upstream SOL for both configurations, the quantities expressed in the upstream location cancel and the final expression takes the form
\begin{equation}
\frac{Q_{\rm \parallel,1}}{Q_{\rm \parallel,2}}\approx\frac{R_{\rm 2}^{\rm t}}{R_{\rm 1}^{\rm t}}\approx \frac{FX_{\rm tor,2}}{FX_{\rm tor,1}},
\end{equation}
i.e. the parallel energy flux in SXD is reduced by the effect of the magnetic topology due to the toroidal flux expansion. The reduced poloidal magnetic field in SXD accounts for the drop of the poloidal energy flux $Q_{\rm pol}=Q_{\parallel}B_x/B$ as
\begin{equation}
\frac{Q_{\rm pol,1}}{Q_{\rm pol,2}}\approx \frac{R_{\rm 2}^{\rm t}}{R_{\rm 1}^{\rm t}} \frac{(\frac{B_x}{B})_{\rm 1}^{\rm t}}{(\frac{B_x}{B})_{\rm 2}^{\rm t}}\approx \frac{B_{x,1}^{\rm t}}{B_{x,2}^{\rm t}} \approx \frac{FX_{\rm tor,2}}{FX_{\rm tor,1}}\frac{FX_{\rm pol,2}}{FX_{\rm pol,1}}
\end{equation}
and the target power load $Q_{\rm t}=Q_{\rm pol}{\rm sin}\beta$ is reduced as
\begin{equation}
\frac{Q_{\rm t,1}}{Q_{\rm t,2}}\approx \frac{R_{\rm 2}^{\rm t}}{R_{\rm 1}^{\rm t}} \frac{(\frac{B_x}{B})_{\rm 1}^{\rm t}}{(\frac{B_x}{B})_{\rm 2}^{\rm t}}\frac{{\rm sin}\beta_1}{{\rm sin}\beta_2} \approx \frac{FX_{\rm tor,2}}{FX_{\rm tor,1}}\frac{FX_{\rm pol,2}}{FX_{\rm pol,1}}\frac{FX_{\rm tilt,2}}{FX_{\rm tilt,1}}=\frac{FX_{2}}{FX_{1}},
\end{equation}
involving also the expansion due to the target tilting.

\end{document}